\documentclass[paper, letterpaper, 11pt, notoc]{JHEP3}
 
\usepackage{graphicx}
\usepackage{amssymb}
\usepackage{amsmath}
\usepackage{latexsym}

\newcommand{\be}{\begin{equation}}
\newcommand{\ee}{\end{equation}}
\newcommand{\bea}{\begin{eqnarray}}
\newcommand{\eea}{\end{eqnarray}}
\newcommand{\comment}[1]{}
\newcommand{\TeV}{~\mathrm{TeV}}
\newcommand{\GeV}{~\mathrm{GeV}}
\newcommand{\Pl}{\mathrm{Pl}}
\newcommand{\lsim}{\!\mathrel{\hbox{\rlap{\lower.55ex \hbox{$\sim$}} \kern-.34em \raise.4ex \hbox{$<$}}}}
\newcommand{\gsim}{\!\mathrel{\hbox{\rlap{\lower.55ex \hbox{$\sim$}} \kern-.34em \raise.4ex \hbox{$>$}}}}
\DeclareMathOperator{\diag}{diag}
\DeclareMathOperator{\tr}{tr}
\DeclareMathOperator{\Tr}{Tr}
 
 \newcommand\openone{\leavevmode\hbox{\small1\normalsize\kern-.33em1}}
\newcommand{\slashed}[1]{ #1 \hspace{-5pt} \slash \hspace{0pt}}
 
 \preprint{HUTP-05/A0002}
 
 \title{The Littlest Higgs in Anti-de Sitter Space}
 
 \author{Jesse Thaler and Itay Yavin \\ Jefferson Physical Laboratory, Harvard University, Cambridge, MA 02138 \\ E-mail:  \email{jthaler@jthaler.net, yavin@fas.harvard.edu}}
 
 \abstract{We implement the $SU(5)/SO(5)$ littlest Higgs theory in a slice of 5D Anti-de Sitter space bounded by a UV brane and an IR brane.  In this model, there is a bulk $SU(5)$ gauge symmetry that is broken to $SO(5)$ on the IR brane, and the Higgs boson is contained in the Goldstones from this breaking.  All of the interactions on the IR brane preserve the global symmetries that protect the Higgs mass, but a radiative potential is generated through loops that stretch to the UV brane where there are explicit $SU(5)$ violating boundary conditions.  Like the original littlest Higgs, this model exhibits collective breaking in that two interactions must be turned on in order to generate a Higgs potential.  In AdS space, however, collective breaking does not appear in coupling constants directly but rather in the choice of UV brane boundary conditions.  We match this AdS construction to the known low energy structure of the littlest Higgs and comment on some of the tensions inherent in the AdS construction.  We calculate the 5D Coleman-Weinberg effective potential for the Higgs and find that collective breaking is manifest.  In a simplified model with only the $SU(2)$ gauge structure and the top quark, the physical Higgs mass can be of order $200 \GeV$ with no considerable fine tuning (25\%).  We sketch a more realistic model involving the entire gauge and fermion structure that also implements $T$-parity, and we comment on the tension between $T$-parity and flavor structure.}
 
\begin{document}

\section{Introduction}

The little Higgs mechanism \cite{Arkani-Hamed:2001nc,Arkani-Hamed:2002pa,Arkani-Hamed:2002qy,Arkani-Hamed:2002qx,Gregoire:2002ra,Low:2002ws,Kaplan:2003uc,Skiba:2003yf} offers a fascinating way to stabilize the electroweak scale.  Like the Georgi-Kaplan composite Higgs \cite{Kaplan:1983fs,Kaplan:1983sm,Georgi:1984af,Georgi:ef,Dugan:1984hq}, the little Higgs is naturally light because it is a pseudo-Goldstone boson of a spontaneously broken approximate global symmetry.  What is novel about the little Higgs mechanism is that it implements collective breaking.  That is, the interactions of the theory are arranged such that turning on any one interaction preserves enough of the global symmetry to protect the Higgs mass.  Therefore, quadratic divergences to the Higgs mass can only appear at mutliloop order, and generically the Higgs boson is two loop-factors lighter than the scale of spontaneous symmetry breaking.  In principle, this allows us to push the cutoff of the standard model as high as $10 \TeV$ without fine-tuning the Higgs mass.  In addition, specific little Higgs models, such as models with a custodial $SU(2)$ symmetry \cite{Chang:2003un,Chang:2003zn} and models which implement $T$-parity \cite{Cheng:2003ju,Cheng:2004yc,Low:2004xc}, automatically satisfy precision electroweak constraints \cite{Hewett:2002px,Csaki:2002qg,Csaki:2003si,Gregoire:2003kr,Kilic:2003mq,Chen:2003fm,Yue:2004xt}.  Therefore, it is reasonable to say that the little Higgs mechanism is a fully realistic theory for perturbative physics up to $10 \TeV$.

Despite the phenomenological successes of little Higgs mechanism, there are questions that one might be interested in answering that cannot be unambiguously calculated in a theory with a $10 \TeV$ cutoff.
In particular, four-fermion operators that generate flavor-changing neutral currents (FCNCs) generically require a cutoff of $\Lambda \sim 1000 \TeV$, so in a theory with a $10 \TeV$ cutoff, we have no natural way of explaining why FCNCs are so suppressed.  Also, in the original little Higgs papers, the Higgs potential is governed by operators with quadratically sensitive couplings, and estimates of these couplings were generated using the Coleman-Weinberg potential \cite{Coleman:jx} and na\"{\i}ve dimensional analysis \cite{Manohar:1983md,Georgi:1986kr}.  While we have no \emph{a priori} reason for not trusting these estimates, there are always $\mathcal{O}(1)$ effects (including sign ambiguities) accompanying UV sensitive physics.  For these reasons, we wish to present a possible UV completion of the little Higgs mechanism in which the UV sensitive physics is finite and calculable.

Here, we will focus on the simplest little Higgs model, the littlest Higgs \cite{Arkani-Hamed:2002qy}.  The littlest Higgs is based on an $SU(5)/SO(5)$ non-linear sigma model, and there are two obvious UV completions one might explore.  The most na\"{i}ve UV completion is a linear sigma model with a Mexican-hat potential, but such a theory cannot address the hierarchy problem because the mass of the linear sigma field is quadratically sensitive to the Planck scale.  Alternatively, the $SU(5) \rightarrow SO(5)$ symmetry breaking pattern could be generated by technicolor-like strong dynamics.   This is the path taken in \cite{Katz:2003sn} based on an $SO(7)$ confining gauge group, and the $SU(5)/SO(5)$ non-linear sigma model arises analogously to the pions of QCD.

Another type of strong dynamics is suggested by the AdS/CFT correspondence \cite{Maldacena:1997re,Gubser:1998bc,Witten:1998qj} and its phenomenological interpretation \cite{Arkani-Hamed:2000ds,Rattazzi:2000hs}.  As noted in \cite{Contino:2003ve}, the Higgs can emerge as a composite state of a strongly coupled 4D quasi-conformal field theory, and while the details of the strong dynamics are difficult to calculate, UV sensitive parameters such as the Higgs potential can be determined in the weakly coupled AdS dual of the quasi-CFT.   In all composite Higgs models, some separation between the confinement scale and the electroweak scale is necessary in order to satisfy precision electroweak constraints.    Recently, the Higgs as a holographic pseudo-Goldstone boson has been incorporated into a realistic model \cite{Agashe:2004rs} where this separation is achieved by a mild fine-tuning between radiative contributions in the fermion sector.  In the littlest Higgs, the Higgs quartic coupling is naturally large and parametrically of order the standard model gauge coupling, so it is particularly interesting to study the composite littlest Higgs in the context of AdS/CFT.  We can imagine a quasi-CFT with an $SU(5)$ global symmetry spontaneously broken to $SO(5)$ at the scale of conformal symmetry breaking, and the $SU(5)/SO(5)$ non-linear sigma model will naturally emerge as composites of the strong dynamics. 

In this paper, we construct a simple AdS$_5$ model \cite{Randall:1999ee,Randall:1999vf} that includes all the major features of the littlest Higgs.    We consider a slice of AdS$_5$ space bounded by a UV brane and an IR brane.    In AdS language, it is easy to see why the Higgs potential will be finite and calculable.  The non-linear sigma field that contains the Higgs doublet lives on the IR brane, and all interactions on the IR brane respect the global symmetries that leave the Higgs massless.  These global symmetries are only broken on the UV brane, so a quantum effective potential for the Higgs is generated through loops that stretch from the IR brane to the UV brane.  Because these loops are non-local in the bulk of AdS, they are manifestly finite.

Even more interesting is what collective breaking looks like in AdS language.  In the original littlest Higgs model, the global $SU(5)$ symmetry that protects the Higgs mass is broken by gauging two subgroups of the $SU(5)$ in such a way that the Higgs is exactly massless unless both gauge couplings are non-zero.   In our AdS model, two subgroups of the bulk $SU(5)$ gauge bosons are given different boundary conditions on the Planck brane, and collective breaking in AdS language is now a statement about collectively choosing these boundary conditions.  Whereas the low-energy Coleman-Weinberg potential involved combinations of coupling constants, the full 5D Coleman-Weinberg potential involves differences of Greens functions in a way that makes collective breaking manifest.

Though the radiatively generated Higgs potential is parametrically of the right form to successfully trigger electroweak symmetry breaking, when we look at the model in numerical detail, we find an interesting tension that is already evident in the low energy phenomenology and which is exacerbated by the AdS construction.  The littlest Higgs contains a top partner $t'$ and electroweak gauge partners $W'$ that cancel quadratic divergences coming from standard model fermion and gauge loops.  The ratio of the $t'$ mass to the $W'$ mass is generically of order $4\lambda_{\mathrm{top}}/\sqrt{g_1^2 + g_2^2}$, where $\lambda_\mathrm{top}$ is the top Yukawa coupling, $g_i$ are $SU(2)_i$ gauge couplings, and the electroweak gauge coupling is determined by $g_{\mathrm{EW}} = g_1 g_2/\sqrt{g_1^2 + g_2^2}$.  For phenomenological reasons, we would like the $t'$ and $W'$ to be nearly degenerate, and therefore we would like to make $g_2 \gg g_1$.  However, in a large $N$ CFT, gauge couplings are infrared free, so we either have to shrink the conformal window in order to enforce a large separation between $g_1$ and $g_2$,  or we have to reckon with dangerously light $W'$ gauge bosons.

In the limit where $g_1 = g_2$, we may be able to live with a light $W'$ if the theory has $T$-parity, and at the end of this paper, we work towards constructing a realistic AdS$_5$ model that implements $T$-parity.   $T$-parity is a $\mathbf{Z}_2$ symmetry under which standard model fields are even but new fields at the TeV scale are odd.  This symmetry forbids standard model dimension six operators from being generated by tree-level exchange of the new particles, so corrections to precision electroweak observables are generically suppressed by a loop factor.  However, we will see a tension between implementing $T$-parity and trying to capitalize on the past successes of AdS$_5$ model building (see \cite{Agashe:2003zs} for a summary).  In particular, there is natural way to generate hierarchies among the Yukawa couplings by putting standard model fermions in the bulk of AdS space with different bulk masses \cite{Grossman:1999ra,Gherghetta:2000qt,Huber:2000ie,Huber:2003tu}, and as an added bonus, this setup naturally leads to suppressed FCNCs.  Unfortunately, the same mechanism that generates Yukawa hierarchies also leads to a hierarchy in the masses of $T$-odd fermion partners, and the lightest $T$-odd fermion is far lighter than the current experimental bound.  This tension between precision electroweak constraints and FCNCs appears to be a problem for any AdS$_5$ model with $T$-parity.  In this light, it may be interesting to look at the AdS implementation of little Higgs theories with a custodial $SU(2)$ symmetry, where the mechanism that protects precision electroweak corrections can coexist with the AdS mechanism of suppressing FCNCs.

Putting aside these detailed model building questions, the littlest Higgs in AdS space is a technically natural extension of the standard model that makes sense up to the Planck scale (or some suitably high UV scale when $g_1 \not= g_2$).  In CFT language, the hierarchy between the Planck scale and the electroweak scale is generated in two steps.  First, the hierarchy between $M_\Pl$ and confinement scale is generated through dimensional transmutation.  Second, the hierarchy between confinement scale and the electroweak scale is guaranteed by collective breaking.  We expect it to be straightforward to generalize this AdS$_5$ construction to other little Higgs theories, allowing us to embed a class of phenomenologically successful models into a UV complete framework.

In the next section, we review the holographic description of spontaneous symmetry breaking and apply it to generic little Higgs theories.  We present the littlest Higgs in AdS space in Section \ref{section:model}, and match this theory to the well-known low energy description in Section \ref{section:Low Energy Matching}.  We comment on the tensions between the low energy theory and a large $N$ CFT description in Section \ref{sec:tensions}.    In Sections \ref{section:collective} and \ref{section:Phenomenology} we calculate the radiatively generated potential for the Higgs doublet and show that there is a wide range of parameters with realistic values for the Higgs mass.  Finally, in Section \ref{sec:tparity} we sketch how to implement $T$-parity in AdS space, and conclude with a preview of future directions in little Higgs model building.

\section{The Little Higgs Mechanism through AdS/CFT}

The little Higgs mechanism is ideally suited for a holographic interpretation because the underlying physics involves symmetry breaking patterns and not the specific mechanism for symmetry breaking.  The generalization of the AdS/CFT conjecture to AdS$_5$ with boundaries \cite{Arkani-Hamed:2000ds,Porrati:2001gx} can only give us limited information about the structure of the CFTs dual to AdS models.  Given an RS1-type model with a warped dimension \cite{Randall:1999ee}, there is a corresponding 4D quasi-conformal field theory coupled to gravity and an elementary sector, and this quasi-CFT confines at the scale of the IR brane.  But in any RS1-type model, we are ignorant about physics on the IR brane at energies near the IR scale because the KK gravitons are strongly coupled there.  The dual description of 5D ignorance is 4D ignorance, so while the AdS/CFT correspondence can tell us a lot about the low-energy bound states of  the dual CFT, we do not have an unambiguous understanding of physics near the confinement scale.

Our goal is to understand the UV sensitive parameters of the littlest Higgs model, and we can arrange the interactions of the theory such that the Higgs potential is insensitive to the details of physics near the confinement scale.  \comment{This may seem a bit counterintuitive, as we are assuming strong dynamics above the confinement scale.}  In 5D language, this feature will be obvious, as the Higgs boson is physically separated in the warped dimension from the fields that radiatively generate the Higgs potential.  In 4D language, the Higgs potential is generated through interactions between the quasi-CFT and the elementary sector and is therefore independent of the details of CFT dynamics.

\FIGURE[t]{
\includegraphics[scale=0.55]{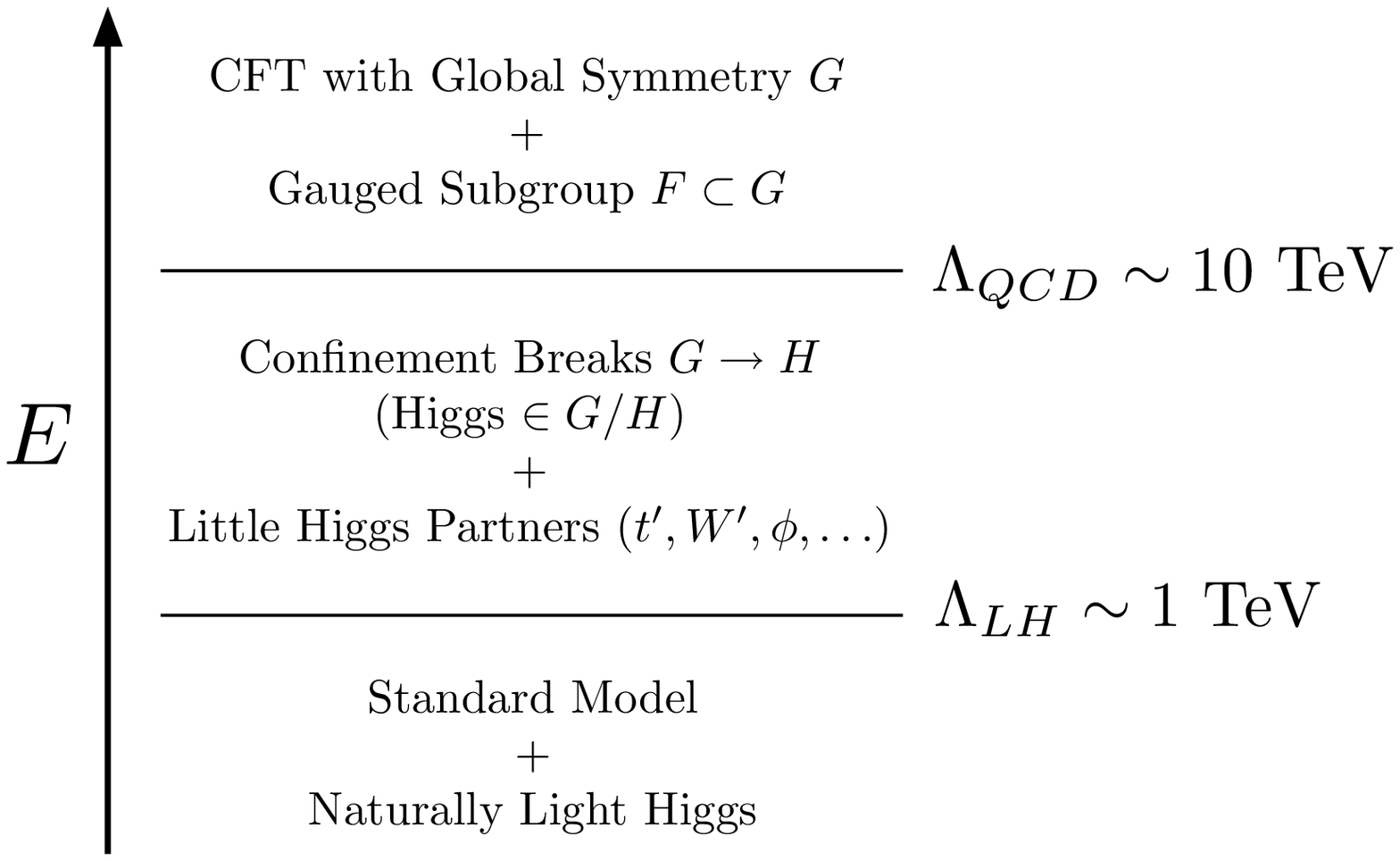}
\caption{Schematic of a generic 4D little Higgs theory. The scale $\Lambda_{LH}$ is the mass of the particles responsible for collective breaking and is roughly of order $f_\pi$, the pion decay constant of the CFT.   The scale $\Lambda_{QCD}$ is where we see any particles not directly involved in collective breaking,  namely the $\rho$ mesons of the CFT.}
\label{cftmodel}
}

%\begin{figure}
%\begin{center}
%\includegraphics[scale=0.60]{cftmodel}
%\end{center}
%\caption{Schematic of a generic 4D little Higgs theory. The scale $\Lambda_{LH}$ is the mass of the particles responsible for collective breaking and is roughly of order $f_\pi$, the pion decay constant of the CFT.   The scale $\Lambda_{QCD}$ is where we see any particles not directly involved in collective breaking,  namely the $\rho$ mesons of the CFT.}
%\label{cftmodel}
%\end{figure}

%\begin{center}
%CFT with Global Symmetry $G$\\
%$+$\\
%Gauge Subgroup $F$
%\end{center}

%\begin{center}
%Confinement Breaks $G \rightarrow H$\\
%(Higgs $\in G/H$)\\
%$+$\\
%Little Higgs Partners ($t', W', \phi, \ldots$)
%\end{center}

Ignoring the fermion sector, all little Higgs models involve a global symmetry $G$ that is spontaneously broken to $H$ at an energy $\Lambda \sim 10 \TeV$.  We will assume that the breaking $G \rightarrow H$ is triggered by QCD-like confinement, and the Goldstone bosons of $G \slash H$ emerge as bound states of the strong dynamics, analogously to the pions of QCD.  (See Figure \ref{cftmodel}.)  A subgroup $F \subset G$ is gauged and contains two copies of the electroweak gauge group $W_i = SU(2)_i \times U(1)_i$ for $i = 1,2$, but $H$ only contains the diagonal electroweak gauge group $W_1 + W_2$.  (In the model we construct in the next section, we only introduce one copy of $U(1)_Y$, so $W_i = SU(2)_i$.)  The Higgs sector is embedded in the pseudo-Goldstone bosons of $G \slash H$ such that the Higgs field transforms as a (diagonal) electroweak doublet.  To implement collective breaking, the structure of the symmetry groups is chosen such that the Higgs is an exact Goldstone boson if only one of the electroweak gauge groups is turned on.

\FIGURE[t]{
\parbox{5in}{\centering
\includegraphics[scale=0.65]{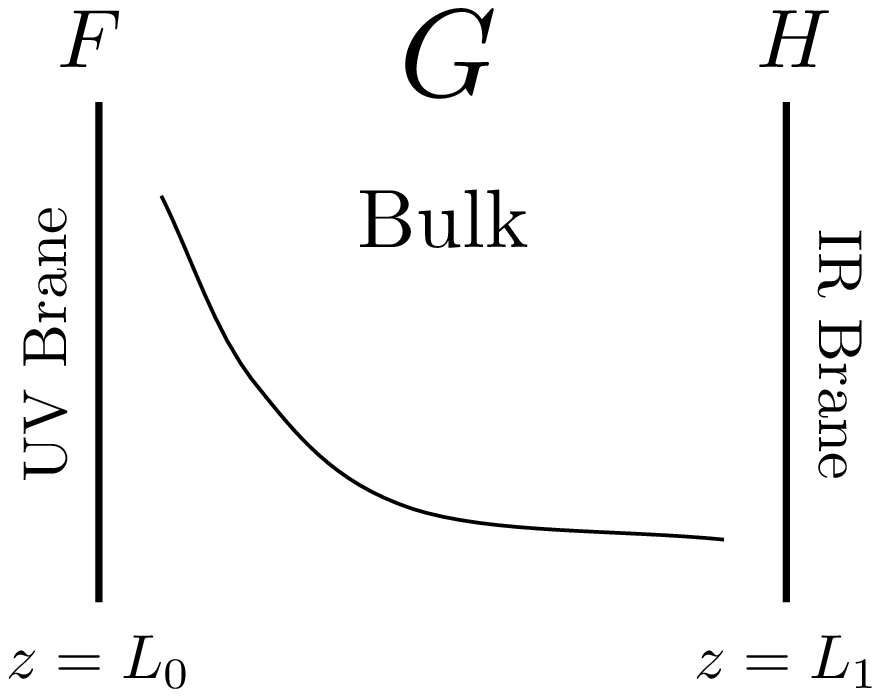}}
\caption{The gauge symmetries in AdS$_5$ for a generic little Higgs theory where there is a $G/H$ non-linear sigma model at low energies with $F \subset G$ gauged.  The reduced gauge symmetries $F$ and $H$ on the UV/IR branes can be accomplished through boundary conditions or linear sigma models.  We will work as closely as possible to boundary condition breaking.}
\label{adsfigure}
}

%\begin{figure}
%\begin{center}
%\includegraphics[scale=0.65]{bulksym}
%\end{center}
%\caption{The gauge symmetries in AdS$_5$ for a generic little Higgs theory where there is a $G/H$ non-linear sigma model at low energies with $F \subset G$ gauged.  The reduced gauge symmetries $F$ and $H$ on the UV/IR branes can be accomplished through boundary conditions or linear sigma models.  We will work as closely as possible to boundary condition breaking.}
%\label{adsfigure}
%\end{figure}

If the strong sector of this theory is a strongly coupled conformal field theory, then there is a simple AdS metaphor for this symmetry breaking pattern.  Consider a slice of AdS$_5$ bounded by a UV brane and a IR brane as in Figure \ref{adsfigure}.  \comment{(We will loosely follow the notation in \cite{Contino:2003ve}.)}  The metric for AdS$_5$ is
\be
\label{eqn:metric}
ds^2 = \frac{1}{(kz)^2} (\eta_{\mu\nu} dx^\mu dx^\nu - dz^2),
\ee
where $k \sim M_\Pl$ is the AdS curvature, and $z$ parametrizes distance in the the warped dimension.  The UV brane sits at $z = L_0 \sim 1/M_\Pl$ and the IR brane sits at $z= L_1 \sim 1/(\mbox{a few} \TeV)$.  It is also convenient to introduce the parameter $\epsilon = L_0/L_1$.  In general, fields that ``live'' on the UV brane correspond to 4D elementary fields that are outside of the CFT, and fields that live on the IR brane correspond to composites of the strong dynamics.  Fields living in the bulk of AdS$_5$ encode information about the operator content of the CFT.  In particular, the global symmetry $G$ of the CFT corresponds to bulk $G$ gauge bosons in AdS$_5$.  Gauging a subgroup $F$ of $G$ in 4D language corresponds to reducing the bulk $G$ gauge symmetry to $F$ on the UV brane.    Spontaneously breaking $G \rightarrow H$ at $\Lambda \sim 10 \TeV$ corresponds to reducing the bulk $G$ gauge symmetry to $H$ on the IR brane.

As shown in \cite{Contino:2003ve}, we could work entirely in the language of boundary conditions in AdS space, imposing Dirichlet boundary conditions on the appropriate modes of the gauge bosons $A_\mu^a$ to reduce the gauge symmetries on the boundaries.  In this case, the uneaten Goldstone modes (including our Higgs boson) correspond to $A_5^a$ zero modes.  For our purposes, we work instead in $A_5=0$ gauge and reduce the gauge symmetry on the IR brane via a linear sigma field $\Phi$ that takes a non-zero vacuum expectation value.  For low energy physics $E < \mbox{a few} \TeV$, the boundary condition language and the linear sigma field language give identical results, and we will go to a limit which is maximally similar to boundary condition breaking.  We prefer the $\Phi$ language because the fluctuations $\Sigma$ about $\langle \Phi \rangle$ \emph{are} the pseudo-Goldstone bosons associated with $G / H$, so the fact that the $\Phi$ field lives on the IR brane makes it intuitively obvious that the pseudo-Goldstone bosons $\Sigma$ are composite states of the strong dynamics.

We still have enough gauge freedom to use boundary conditions to reduce the gauge symmetry on the UV brane to $F$.  As we will see explicitly, collective breaking now looks like a choice between Neumann and Dirichlet boundary conditions for the electroweak groups $W_1$ and $W_2$.  If $W_1$ has Neumann boundary conditions but $W_2$ has Dirichlet boundary conditions (\emph{i.e.}\ only $W_1$ is gauged), then the Higgs contained in $\Sigma$ is still an exact Goldstone.  Only when all of $F$ has Neumann boundary condition does the Higgs pick up a mass.   This is because $\Phi$ transforms as some representation of the bulk gauged $G$, and $G$-violating corrections to the $\Phi$ potential can only come through loops involving gauge bosons that stretch from the IR brane (where $G$ is a good symmetry before $\Phi$ takes its vev) to the UV brane (where $G$ is explicitly broken by boundary conditions).  Because these loops are non-local in the warped dimension, their contribution to the Goldstone potentials are finite and calculable.

To complete the little Higgs mechanism we need to introduce fermions.  Because the Higgs lives on the IR brane, standard model fermions must have some overlap with the IR brane.  Standard model fermions do not come in complete $G$ multiplets, however, and if we simply included $G$-violating Yukawa couplings with the $\Phi$ field on the IR brane, then the quantum effective potential for the $\Sigma$ field would be sensitive to the cutoff on the IR brane, namely $\mbox{a few} \TeV$.  While this is not a problem from a model-building perspective, our aim was to have control over UV sensitive parameters in little Higgs models.  Therefore, as for the gauge sector, we only want to explicitly break $G$ on the UV brane, so we introduce bulk fermions in complete $G$ multiplets.  These bulk fermions couple to the $\Phi$ field on the IR brane and have explicit $G$-violating (but $F$-preserving) boundary conditions on the UV brane.  Like the gauge sector, these boundary conditions are chosen collectively such that a mass for the Higgs is only generated when there are pairs of Dirichlet boundary conditions.

While it appears that collective breaking involves a binary choice between Neumann and Dirichlet boundary conditions, we could of course break $G$ by introducing $G$-violating UV brane kinetic terms for the gauge bosons and fermions.   In our model, we in fact need to do this because simple boundary conditions are not enough to generate different gauge couplings for $W_1$ and $W_2$.  Clearly, if we introduced a boundary mass term proportional to $m$, then Neumann and Direchlet boundary conditions are just the $m \rightarrow 0$ and $m \rightarrow \infty$ limits, respectively.  As we will see when we write down the expression for the Higgs potential, the boundary condition language makes it easy to see how the little Higgs mechanism works in anti-de Sitter space, but the central idea is applicable for general boundary kinetic terms.  We use the shorthand $(\pm, \pm)$ to indicate boundary conditions on the UV/IR branes, where $+$ indicates Neumann (or modified Neumann) boundary conditions, and $-$ indicates Dirichlet boundary conditions.

\section{An AdS Implementation of the Littlest Higgs}

\label{section:model}

%\begin{figure}
%\begin{center}
%\includegraphics[scale=0.50]{choosensym}
%\end{center}
%\caption{Gauge structure of the littlest Higgs in AdS$_5$.  }
%\label{figure:littleinads}
%\end{figure}

\FIGURE{
\parbox{5in}{\centering
\includegraphics[scale=0.45]{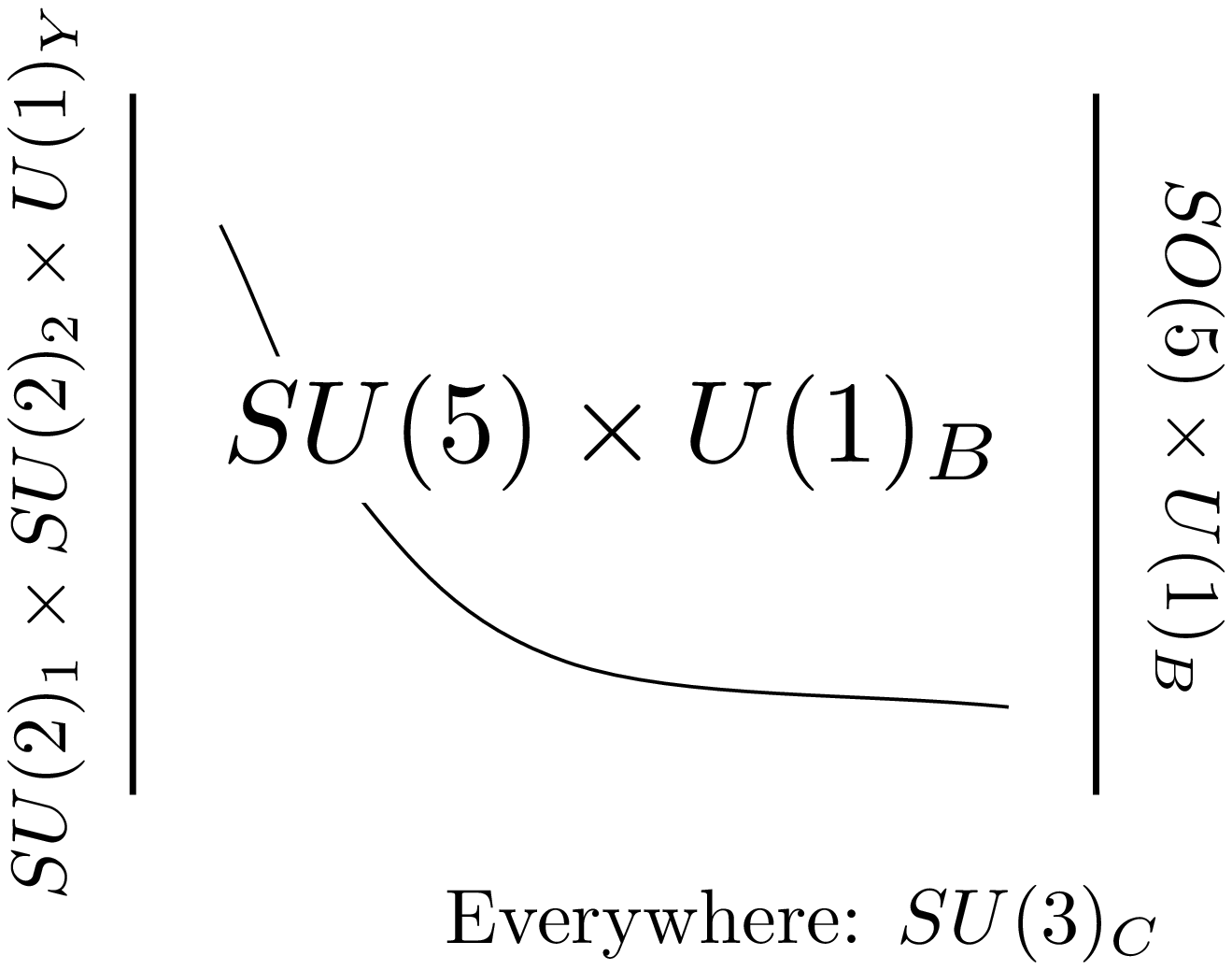}}
\caption{Gauge structure of the littlest Higgs in AdS$_5$.  }
\label{figure:littleinads}
}

The littlest Higgs \cite{Arkani-Hamed:2002qy} is based on an $SU(5)/SO(5)$ non-linear sigma model.  In this section, we construct an AdS$_5$ model that captures the essential features of the littlest Higgs involving only the top and gauge sectors. This model roughly shares the 4D fermion content of \cite{Katz:2003sn}.  Unlike the original littlest Higgs, we do not implement collective breaking for $U(1)_Y$ in order to avoid issues arising from mixing between heavy and light $U(1)$ gauge bosons, but because the hypercharge gauge coupling is small, this will not change the Higgs mass appreciably.  In the calculations in the following sections, we simply ignore the $U(1)_Y$ contribution to the Higgs mass and quartic coupling.

In the language of the previous section, we will take
\be
\begin{array}{ccccc}
G &=&SU(3)_C& \times& SU(5)  \times U(1)_B  \\
F &=&SU(3)_C& \times& SU(2)_1 \times SU(2)_2  \times U(1)_Y \\
H&=&SU(3)_C& \times& SO(5)  \times U(1)_B
\end{array}
\ee
where $SU(3)_C$ is color $SU(3)$, and $U(1)_B$ has nothing to do with baryon number.   (See Figure \ref{figure:littleinads}.)  The $SU(2)$ generators of $F$ are imbedded in the $SU(5)$ generators as
\be
Q_1^a =\left(\begin{array}{cc}\sigma^{a}/2 & \quad \\ \quad&\quad \end{array}\right), \qquad Q_2^a = \left(\begin{array}{cc} \quad&\quad  \\ \quad &-\sigma^{*a}/2\end{array}\right).
\ee
There is a $U(1)_A$ in $SU(5)$ generated by $A = \diag(1,1,0,-1,-1)/2$, and $U(1)_Y$ is defined by $Y = A + B$, where $B$ is the $U(1)_B$ generator.  For the purposes of calculating the radiative Higgs potential, we need only introduce the bulk gauge coupling $g_5$ for $SU(5)$.  We also introduce a Planck brane boundary gauge kinetic terms for the $SU(2)_i$ subgroups proportional to $z_i$.  Propagators for gauge fields with boundary conditions and Planck brane kinetic terms are given in Appendix \ref{conventions}.  The low energy gauge couplings $g_i$ for the $SU(2)_i$ can be extracted by integrating out the bulk of AdS:
\be
\label{eqn:LowEnergyGaugeCoupling}
\frac{1}{g_i^2} = \frac{1}{g^2_5 L_0}\left(\log \frac{L_1}{L_0} + z_i \right) = \frac{1}{g_\rho^2}(-\log \epsilon + z_i),
\ee
where we have introduced $g_\rho = g_5 \sqrt{L_0}$.  $g_\rho$ is the AdS perturbative expansion parameter that has the holographic interpretation $g_\rho \sim 4\pi/ \sqrt{N}$ in a large $N$ CFT \cite{Arkani-Hamed:2000ds}.

% (\emph{i.e.}\ it is a $(\mathbf{5} \otimes \mathbf{5})_\mathrm{sym} = \mathbf{15}$)

The field $\Phi$ on the IR brane is a symmetric tensor that transforms as
\be
\Phi \rightarrow V \Phi V^T
\ee
under $SU(5)$, and $\Phi$ is a singlet under $SU(3)_C \times U(1)_B$.  $\Phi$ takes a vacuum expectation value
\be
\langle \Phi \rangle = \Sigma_0 \equiv \left(
\begin{array}{ccc} &  & \openone \\ & 1 &  \\ \openone &  &
\end{array}\right),
\ee
breaking $SU(5) \rightarrow SO(5)$ on the IR brane.  The unbroken $SO(5)$ generators $T_a$ and the broken $SU(5)/SO(5)$ generators $X_a$ satisfy
\be
T_a \Sigma_0 + \Sigma_0 T_a^T = 0, \qquad X_a \Sigma_0 - \Sigma_0 X_a^T = 0.
\ee
Performing a broken $SU(5)$ transformation on the vacuum, we can parametrize fluctuations about the $\Phi$ background as
\be
\Sigma = e^{i\Pi/f_5} \Sigma_0 e^{i\Pi^T/f_5} = e^{2i\Pi/f_5} \Sigma_0,
\ee
where $\Pi = \pi^a X_a$ is the Goldstone matrix and $f_5$ is related the 4D Goldstone decay constant $f_4$ by $f_4 = \epsilon f_5$.  As we will see, $f_4$ has nothing to do with what a low energy observer would call the pion decay constant, and we will actually go to the limit $f_4 \rightarrow \infty$.  Three of the Goldstone bosons are eaten by the Higgsing of $SU(2)_1 \times SU(2)_2$ down to $SU(2)_{EW}$, giving rise to the gauge boson partner $W'$.  The remainder of the $\Pi$ matrix can be written as
\be
\label{eqn:pifield}
\Pi = \left(\begin{array}{ccc} (\eta/\sqrt{20}) \openone  & h^\dagger/ \sqrt{2} & \phi^\dagger \\h/\sqrt{2} & -4\eta / \sqrt{20} & h^* /\sqrt{2} \\\phi & h^T / \sqrt{2} & (\eta/ \sqrt{20}) \openone \end{array}\right),
\ee
where $\eta$ is a real neutral field, $h = (h^+, h^0)$ is the Higgs doublet, and $\phi$ is a complex symmetric two by two matrix that transforms as a charge 1 electroweak triplet.  

On the IR brane, the leading lagrangian for $\Sigma$ is the gauged non-linear sigma model
\be
\label{eqn:SigmaKinetic}
\mathcal{L}_\Sigma = \sqrt{-g_\mathrm{ind}} \delta(z-L_1)g_{\mathrm{ind}}^{\mu \nu} \frac{f_5^2}{8} \tr (D_\mu \Sigma) ^\dagger (D_\nu \Sigma), \qquad D_\mu \Sigma= \partial_\mu \Sigma - i A_\mu \Sigma - i \Sigma A_\mu^T,
\ee
where $A_\mu = A^a_\mu T_a + A^b_\mu X_b$ are the 5D $SU(5)$ gauge bosons.  In order to give the $\eta$ field a mass and remove it from the spectrum, we introduce an explicit $SO(5)$ violating plaquette operator on the IR brane which does not substantially affect the Higgs potential.  Expanding $\Sigma$ in the Goldstone fields, there is a linear coupling between $h$ and the KK gauge bosons which induces wavefunction renormalization on $h$.  Integrating out the heavy KK states, the kinetic lagrangian for $h$ is
\be
\mathcal{L}_{h} = Z_h |D_\mu h |^2, \qquad Z_h = \left(1 +  \frac{f_5^2 g_5^2 G_{br}}{2}  \right)^{-1},
\ee
where the covariant derivative now only includes electroweak gauge fields, and $G_{br} = \hat{G}(0;L_1,L_1) = L_0 / 2$ is the zero momentum boundary to boundary rescaled gauge boson propagator for modes with $(-,+)$ boundary conditions.

In this simplified littlest Higgs model, we only consider fermionic contributions to the Higgs mass from the top sector.  We introduce two bulk 5D (Dirac) fermions $\mathbf{Q}$ and $\mathbf{Q^c}$.  Under KK decomposition, $\mathbf{Q}$ and $\mathbf{Q^c}$ have upper (left-handed) and lower (right-handed) Weyl components:
\be
\mathbf{Q} = \left(\begin{array}{c}Q \\\overline{Q}' \end{array}\right), \qquad \mathbf{Q^c} = \left(\begin{array}{c}Q'^c \\\overline{Q}^c\end{array}\right).
\ee
In order for the 4D masses of the KK fermion tower to have real masses, we must impose Dirichlet boundary conditions on either the upper or lower component on each of the AdS boundaries \cite{Grossman:1999ra}.  However, because we do not impose orbifold symmetry on AdS space, the choice of which mode vanishes is independent on each boundary.  We choose the upper component $Q$ of $\mathbf{Q}$ and the lower component $Q^c$ of $\mathbf{Q^c}$ to have non-vanishing boundary conditions on the IR brane.  While this choice is arbitrary, it will simplify the calculation of the fermionic contribution to the $\Sigma$ effective potential.

The components $Q$ and $Q^c$ transform under $G$ as:
\be
\begin{array}{c|ccc}
&SU(3)_C & SU(5) & U(1)_B\\
\hline
Q & \mathbf{3} & \mathbf{5} & +2/3\\
Q^c & \mathbf{\bar{3}} & \mathbf{5} & -2/3
\end{array}
\ee
Expanding $Q$ and $Q^c$ as $SU(5)$ multiplets:
\be
Q = \left(\begin{array}{c}\tilde{p} \\\tilde{t} \\ q \end{array}\right), \qquad Q^c = \left(\begin{array}{c}\tilde{q}^c \\ t^c \\ \tilde{p}^c\end{array}\right).
\ee
Under $F$, these fields transform as:
\be
\label{eqn:FermionTable}
\begin{array}{c|cccc|c}
&SU(3)_C & SU(2)_1 & SU(2)_2 & U(1)_Y& B.C.\\
\hline
q & \mathbf{3} & - & \mathbf{2} & +1/6 &(+,+)\\
t^c & \mathbf{\bar{3}} & - & - & -2/3 &(+,+)\\
\hline
\tilde{t} & \mathbf{3} & - & - & +2/3 &(-,+) \\
\tilde{q}^c & \mathbf{\bar{3}} & \mathbf{2} &-& - 1/6&(-,+) \\
\tilde{p}&\mathbf{3}& \mathbf{2}&-&+7/6&(-,+)\\
\tilde{p}^c & \mathbf{\bar{3}}& - & \mathbf{2} & -7/6&(-,+)
\end{array}
\ee
After $SU(2)^2$ is Higgsed to the electroweak $SU(2)$, the $q$ and $t^c$ fields have the right quantum numbers to be the standard model third generation quark doublet and top singlet.  In order for $q$ ($t^c$) to have zero modes, we take them to have Neumann boundary conditions on the UV brane, and thus the associated lower (upper) component modes have Dirichlet boundary conditions there.  The tilded fields have Dirichlet boundary conditions on the UV brane, so the first KK mode is massive.  The $\tilde{t}$ and $\tilde{q}^c$ fields are roughly the $t'$ and $q'$ fermionic partners responsible for cutting off the divergent top contribution to the Higgs mass, but the $\tilde{p}$ and $\tilde{p}^c$ are spectators that serve merely to fill out the $SU(5)$ representation.\footnote{Though $\tilde{p}$ and $\tilde{p}^c$ have the right quantum numbers such that their zero modes could become one massive vector fermion, we will have more flexibility to choose different bulk masses for $\mathbf{Q}$ and $\mathbf{Q^c}$  if we remove the $\tilde{p}$ and $\tilde{p}^c$ zero modes.}

As is standard in AdS model building, we give $\mathbf{Q}$ and $\mathbf{Q^c}$ bulk masses $\nu k$ and $\nu^c k$.  The values of $\nu$ and $\nu^c$ affect the spectrum and wavefunctions of the KK tower.  In CFT language, the values of $\nu$ and $\nu^c$ adjust the anomalous dimension of operators corresponding to the zero mode of the relevant fermions \cite{Contino:2004vy}:
\be
\gamma = |\nu - 1/2| - 1, \qquad \gamma^c = |\nu^c + 1/2| - 1.
\ee
(Unfortunately, we use opposite conventions from \cite{Contino:2004vy} for the signs of $\nu$ and $\nu^c$.)  In particular, in order to generate a large enough top Yukawa coupling, we need the top doublet and singlet to have anomalous dimension close to or less than zero.  Though we will not discuss precision electroweak tests in this paper, we note that because these anomalous dimensions are indictations of couplings between the fermions with the CFT, deviations from standard model predictions for $Z \rightarrow b \bar{b}$, $S$, $T$, \emph{etc.}\ are expected to be smaller the closer $\gamma$ and $\gamma^c$ are to zero.  (See \cite{Agashe:2003zs, Agashe:2004rs} for details.)  In Model 3 presented in Section \ref{section:Phenomenology}, $\nu = -\nu^c \sim 1/2$ which at first glance seems to suggest large deviations from precision electroweak measurements.  However, in that model, the CFT resonances appear at around $10 \TeV$, so while the mixing with CFT states may be large, the suppression scale for dangerous dimension six operators is much higher.

To generate the top Yukawa coupling, we introduce a Yukawa interaction on the IR brane:
\be
\label{eqn:yukawainteraction}
\mathcal{L}_t = \sqrt{-g_\mathrm{ind}} \delta(z-L_1)  \left( \lambda Q \Sigma^\dagger Q^c + h.c. \right),
\ee
where $\lambda$ is $\mathcal{O}(1)$.\footnote{This is dimensionally correct, because $Q$ and $Q^c$ are components of 5D fermions and have mass dimension $2$.}  This interaction does not break any of the global symmetries acting on $\Sigma$. In fact, ignoring the gauge sector, there is an enhanced $SU(5)_L \times SU(5)_R$ global symmetry acting on $\Sigma$:
\be
\Sigma \rightarrow L \Sigma R^T, \qquad Q \rightarrow LQ, \qquad Q^c \rightarrow RQ^c,
\ee
and \emph{both} of these $SU(5)$s must be broken in order for the Higgs to get a radiative mass from the fermion sector.  Indeed, the UV brane boundary conditions on $\mathbf{Q}$ break $SU(5)_L$ and the UV brane boundary conditions on $\mathbf{Q^c}$ break $SU(5)_R$.   Equivalently, at energies below the mass of the fermion KK modes, all the symmetries that protect the mass of the Higgs boson are broken, but when we see the KK modes of the tilded fields, we restore the $SU(5)_L \times SU(5)_R$ global symmetry of the fermion sector.  In order for the tilded fields to naturally cancel the quadratically divergent top loop, at least one of the top partners must have mass no greater than around $2 \TeV$ \cite{Arkani-Hamed:2002qy}.

Expanding around $\Sigma = \Sigma_0$, we see that the Yukawa interaction in equation (\ref{eqn:yukawainteraction}) generates kinetic mixing between the tilded and untilded fields.  The overlap of the zero modes of $q$ and $t^c$ with the IR brane are
\be
q:  \; f_L(\nu) =  \frac{1}{L_1^{1/2}}\sqrt{\frac{1+2\nu}{1-\epsilon^{1+ 2\nu}}}, \qquad t^c: \;   f_R(\nu^c) =  \frac{1}{L_1^{1/2}}\sqrt{\frac{1-2\nu^c}{1-\epsilon^{1- 2\nu^c}}}.
\ee
Mixing between the zero mode of $q$ ($t^c$) and the KK modes of $\tilde{q}^c$  ($\tilde{t}$) induce the following wavefunction renormalizations:
\be
Z_q = 1 + \lambda^2 f_L(\nu)^2 G_{\tilde{q}^c}/\epsilon, \qquad
Z_{t^c} = 1 + \lambda^2 f_R(\nu^c)^2 G_{\tilde{t}}/\epsilon,
\ee
where $G_{\tilde{q}^c} = \hat{G}^c(0; L_1, L_1;\nu^c)$ and $G_{\tilde{t}} = \hat{G}(0; L_1, L_1;\nu)$ are zero momentum boundary to boundary rescaled fermion propagators for modes with $(-, +)$ boundary conditions. After integrating out the heavy modes, the low energy Yukawa interaction in terms of canonically normalized fields is:
\be
\label{eqn:TopYukawa}
\mathcal{L}_t = \lambda_{\rm top} \left(q h t^c + h.c.\right), \qquad \lambda_{\rm top} =  \frac{\lambda f_L(\nu) f_R(\nu^c)\sqrt{2}}{f_4 \sqrt{Z_h Z_q Z_{t^c}}},
\ee
where using Appendix A, we have
\be
\label{eqn:FieldStrengthNorm2}
Z_h = \left(1 + \frac{f_5^2 g_5^2 L_0}{4} \right)^{-1}, \qquad
Z_q = 1 + \left(\lambda \frac{f_L(\nu)}{f_L(\nu^c)}\right)^2, \qquad
Z_{t^c} = 1 + \left(\lambda \frac{f_R(\nu^c)}{f_R(\nu)}\right)^2.
\ee
In the next section, we will show how to understand this expression in terms of the $t'$ and $q'$ fields of the littlest Higgs.

\section{Matching to the Low Energy Theory}
\label{section:Low Energy Matching}

The low energy phenomenology of the littlest Higgs is governed by the pion decay constant $f_\pi$, which sets the mass scale for the gauge boson partner $W'$, the fermionic partners $t'$ and $q'$, and ultimately the electroweak scale.  In our AdS construction, however, we appear to have two independent mass scales, $f_4$ and $1/L_1$, and it is not clear which combination we should call $f_\pi$.   From the CFT point of view, we want the $SU(5)/SO(5)$ non-linear sigma field to arise directly from confinement and not from a composite linear sigma field.  In other words, we would like to decouple all information about the $\Phi$ field that lived on the IR brane, and the way to do this is to send $f_4 \rightarrow \infty$.  In AdS langauge, the $f_4 \rightarrow \infty$ limit is morally equivalent to describing $\Sigma$ as the zero mode of $A_5$, though by leaving $f_4$ as a finite parameter, we are able to go to the computationally simpler $A_5 = 0$ gauge.

We notice immediately that all reference to $f_4$ vanishes in this limit.  The top Yukawa coupling in equation (\ref{eqn:TopYukawa}) involves the combination $f_4 \sqrt{Z_h}$, and in this limit,
\be
\label{eqn:hubris}
f_4 \sqrt{Z_h} \rightarrow \frac{2 \epsilon}{g_5 \sqrt{L_0}} = \frac{2}{g_\rho L_1} \equiv f_\pi, \qquad \lambda_{\rm top} \rightarrow \frac{\lambda f_L(\nu) f_R(\nu^c)\sqrt{2}}{f_\pi\sqrt{ Z_q Z_{t^c}}},
\ee
where we have cavalierly defined the pion decay constant $f_\pi$.  To justify this choice, we need to match to the gauge sector.   At large $f_4$, the $\Sigma$ kinetic term in equation (\ref{eqn:SigmaKinetic}) enforces vanishing IR boundary conditions on the gauge bosons corresponding to the subgroup $G/H$.  Gauge bosons contained in $F \cap G/H$ are the littlest Higgs $W'$ modes.  If we take $g_1 = g_2$ ($z_1 = z_2$), then using Appendix \ref{conventions} the mass of the lightest KK mode of a $(+, -)$ gauge boson is roughly $m^2_{W'} = 2/L_1^2(-\log \epsilon + z)$.  For arbitrary Planck brane gauge kinetic terms, the $W'$ has approximate mass
\be
m^2_{W'} \sim \frac{1}{L_1^2}\left(\frac{1}{-\log\epsilon + z_1} + \frac{1}{-\log\epsilon + z_2} \right)  = \frac{g_1^2+g_2^2}{g_\rho^2 L_1^2 }.
\ee
where we have used equation (\ref{eqn:LowEnergyGaugeCoupling}) in the last step.  In the littlest Higgs, the mass of the $W'$ is \cite{Arkani-Hamed:2002qy}
\be
m_{W'} = \frac{g' f_\pi}{2}, \qquad g' = \sqrt{g_1^2 + g_2^2}.
\ee
Therefore, to match the low energy theory, we should identify $f_\pi$ with
\be
\label{eqn:gaugefpi}
f_\pi \equiv \frac{2}{g_\rho L_1},
\ee
in agreement with equation (\ref{eqn:hubris}).  The mass of the lightest KK mode of a $(+,+)$ or $(-,+)$ gauge boson is roughly $m_\rho \sim 3\pi/4 L_1$, which sets the scale for the $\rho$-like resonances of the CFT.  In terms of $f_\pi$ and $g_\rho \sim 4 \pi/\sqrt{N}$:
\be
f_\pi \sim \frac{m_\rho}{g_\rho} \sim \frac{m_\rho}{4\pi}\sqrt{N},
\ee
which reproduces the expected $\sqrt{N}$ scaling of $f_\pi$ in strongly coupled theories \cite{'tHooft:1973jz}.   Because we want a large separation between the scale of new physics $m_\rho$ and the pion decay constant $f_\pi$, we would like to take $N$ small ($g_\rho$ large).  For reference, the electroweak gauge coupling $g_{EW}$ in terms of $g_1$ and $g_2$ is
\be
g_{EW} = \frac{g_1 g_2}{\sqrt{g_1^2+g_2^2}}.
\ee

Now that we understand how to identify $f_\pi$ in the gauge sector, we want to know whether this choice is consistent with the fermion sector.  Using the fermion content of \cite{Katz:2003sn} and a notation suggestive of equation (\ref{eqn:FermionTable}), the low energy littlest Higgs fermion sector has the lagrangian
\be
\mathcal{L}_t = \lambda_1 (q h t^c  + \alpha_q f_\pi  q \tilde{q}^c + \alpha_t f_\pi \tilde{t} t^c)  + \lambda_2 f_\pi \tilde{q} \tilde{q}^c + \lambda_3 f_\pi \tilde{t} \tilde{t}^c,
\ee
where $\alpha_i$ are $\mathcal{O}(1)$ parameters inserted for reasons that will become clear shortly.\footnote{ From the low energy point of view, $\alpha_q$ and $\alpha_t$ account for any effects from wave function renormalization.}  The fields $q$, $\tilde{t}$, $t^c$, and $\tilde{q}^c$ are familar from equation (\ref{eqn:FermionTable}), and the interaction proportional to $\lambda_1$ would arise from the leading expansion of equation (\ref{eqn:yukawainteraction}).  Because  $\tilde{q}^c$ and $\tilde{t}$ are $(-,+)$ modes, they do not have zero modes, and $\tilde{q}$ and $\tilde{t}^c$  are the components of $Q'^c$ and $Q'$ that pair with $\tilde{q}^c$ and $\tilde{t}$ to form Dirac masses.  We see that a linear combination of $q$ and $\tilde{q}$ marries with $\tilde{q}^c$ to become the $q'$, and a linear combination of $t^c$ and $\tilde{t}^c$ marries with $\tilde{t}$ to become the $t'$.   The partners $q'$ and $t'$ cut off fermionic quadratic divergences in the Higgs potential, and their masses are
\be
m_{q'} = f_\pi \sqrt{\alpha_q^2 \lambda_1^2 + \lambda_2^2}, \qquad m_{t'} = f_\pi \sqrt{\alpha_t^2 \lambda_1^2 + \lambda_3^2}.
\ee
After integrating out $t'$ and $q'$, the low energy Yukawa coupling is
\be
\label{eqn:lowenergytop}
\lambda_{\rm top} = \frac{\lambda_1}{\sqrt{1+\alpha_q^2 \lambda_1^2 /\lambda_2^2}\sqrt{1 + \alpha_t^2 \lambda_1^2/ \lambda_3^2}}.
\ee

This form of the Yukawa coupling is very suggestive of equation (\ref{eqn:TopYukawa}).  We can match $\lambda_i$ to parameters of the AdS theory by expanding equation (\ref{eqn:yukawainteraction}) in KK modes and canonically normalizing just the Higgs doublet:
\be
\nonumber
\lambda_1 \rightarrow \frac{\lambda f_L f_R \sqrt{2}}{f_4\sqrt{Z_h}}, \qquad \alpha_q \lambda_1 f_\pi \rightarrow \lambda  f_L f^{(1)}_R, \qquad \alpha_t \lambda_1 f_\pi  \rightarrow \lambda f^{(1)}_L f_R,
\ee
\be
\label{eqn:lambdamapping}
\lambda_2 f_\pi \rightarrow m^{(1)}_R, \qquad \lambda_3 f_\pi \rightarrow m^{(1)}_L.
\ee
Here $m^{(1)}_L(\nu)$ and $f_L^{(1)}(\nu)$ are the mass and IR brane overlap of the lightest KK mode of a $(-,+)$ upper component fermion with bulk mass $\nu k$, and similarly for $m^{(1)}_R(\nu^c)$ and $f_R^{(1)}(\nu^c)$.   Note that $f_L$ and $f_R$ have dimensions of $\sqrt{\rm mass}$ so these expressions match dimensionally.  The ratio $\alpha_q/ \alpha_t$ is clearly necessary to account for the fact that $\lambda f_L f^{(1)}_R$ need not equal $\lambda f^{(1)}_L f_R$.  
We see that all of the $\lambda_i$'s act as spurions for the soft breaking of $SU(5)$.  In particular, $\lambda_2$ and $\lambda_3$ are proportional to the masses of $(-,+)$ modes, and if $SU(5)$ were restored on the UV brane, $\lambda_2$ and $\lambda_3$ would go to zero as all the $(-,+)$ modes would become $(+,+)$ zero modes.

We can write the expressions for $\lambda_{\rm top}$ in terms of boundary to boundary 5D propagators.  On the IR brane, $(-,+)$ bulk fields look like tower of massive states with different overlaps:
\be
\hat{G}(p;L_1,L_1)/\epsilon \sim \sum_i  \frac{{f^{(i)}}^2}{p^2 + {m^{(i)}}^2}.
\ee
(We have analytically continued to momentum space.)  As long as there is a large mass separation between the first state the second state, we can write approximately
\be
\hat{G}(0; L_1,L_1)/\epsilon \sim \frac{{f^{(1)}}^2}{{m^{(1)}}^2}.
\ee
From the mapping in equation (\ref{eqn:lambdamapping}),
\be
1+ \alpha_q^2 \lambda_1^2 / \lambda_2^2 \rightarrow Z_q, \qquad 1+ \alpha_t^2 \lambda_1^2 / \lambda_3^2 \rightarrow Z_{t^c},
\ee
and we see that the Yukawa couplings in equations (\ref{eqn:TopYukawa}) and (\ref{eqn:lowenergytop}) match beautifully.

We can use equation (\ref{eqn:lambdamapping}) to identify the low energy $f_\pi$:
\be
f_\pi  = \frac{f_L^{(1)}}{\alpha_t f_L \sqrt{2}}f_4 \sqrt{Z_h} =  \frac{f_R^{(1)}}{\alpha_q f_R \sqrt{2}}f_4 \sqrt{Z_h}.
\ee
This matches to equation (\ref{eqn:hubris}) as long as we define $\alpha_t$ and $\alpha_q$ appropriately. Of course, these were unknown coefficients in the low energy theory anyway so it is not surprising that their value depends on the UV completion.  More importantly, once we fix what we mean by $f_\pi$ from the gauge sector, there is an unambiguous translation to the fermion sector.  Note that when $\nu = \nu_c = 0$, $f_i^{(1)}/f_i \sim \sqrt{2}$, and $\alpha_i \sim 1$.

\section{Tension with the High Energy Theory}
\label{sec:tensions}

Though we have successfully matched our AdS construction with low energy observables,  there are relationships between the parameters of the littlest Higgs which although harmless from the low energy point of view, cause some tension once a UV completion in the form of a large $N$ CFT is chosen, and we will see that in order to address these issues, we need to shrink the conformal window by taking $1/L_0$ to be smaller than $M_\Pl$.  The UV scale can still be quite high, however, so in this sense we can still claim a viable UV completion of the littlest Higgs.

At the end of the day, the radiatively generated Higgs potential will take the form
\be
V(h) = -m_h^2 h^\dagger h + \lambda_h (h^\dagger h)^2, \qquad v_{EW} = \sqrt{\frac{-m_h^2}{\lambda_h}}, \qquad m_{h_0} = \sqrt{-2m_h^2},
\ee
where $v_{EW} \sim 246 \GeV$ is the electoweak scale and $m_{h_0}$ is the mass of the physical Higgs boson.  Because $\lambda_{\rm top}$ is numerically larger than $g_{EW}$, the dominant contribution to the Higgs potential will come from top loops.  From the low energy Coleman-Weinberg potential, there is the logarithmic contribution to the Higgs doublet mass from the fermion sector \cite{Katz:2003sn},
\be
\label{eqn:Fermion Log correction}
\delta_{\rm fermion} m_h^2 = -\frac{3\lambda_{\rm top}^2 }{8\pi^2} \frac{m_{q'}^2 m_{t'}^2 }{m_{q'}^2-m_{t'}^2} \log \frac{m_{q'}^2}{m_{t'}^2} \rightarrow -\frac{3\lambda_{\rm top}^2}{8\pi^2} m_{t'}^2 ,
\ee
where in the last step, we have taken the special case $m_{t'} = m_{q'}$, which minimizes the contribution to the Higgs doublet mass for fixed $\lambda_{\rm top}$.  (We are assuming $\alpha_t = \alpha_q = 1$ for simplicity.)  In this limit, $m_{t'} = 2\sqrt{2} \lambda_{\rm top} f_{\pi}$,\footnote{When $\alpha_i \not= 1$, the mass of the $t'$ can be lighter than $2\sqrt{2} \lambda_{\rm top} f_{\pi}$ even in the limit $m_{t'} = m_{q'}$.}
and the contribution to the Higgs doublet mass is bounded by
\be
\delta_{\rm fermion} m_h^2 = -\frac{3\lambda_{\rm top}^4 f_\pi^2}{\pi^2}.
\ee
The physical Higgs will be roughly a factor of $\lambda_{\rm top}/4$ lighter than the $t'$, and to the extent that $\lambda_{\rm top}$ is small, we will have a light Higgs boson.  Of course, $\lambda_{\rm top} \sim 1$, so numerically there is not much of a separation of scales.  While it may not be a total disaster if $m_{h_0} \sim m_{t'}$, the mass of the $t'$ is set by $f_\pi$ which in turn sets the mass of the $W'$:
\be
m_{W'} = \frac{g' f}{2} = m_{t'}\frac{g'}{4\sqrt{2} \lambda_{\rm top}}.
\ee
If $g_1 = g_2$, then $g' = 2 g_{EW}$, and $m_{W'}$ would generically be $\emph{lighter}$ than the Higgs boson.

The obvious way to relieve this tension is to raise $g'$ by increasing $g_2$, but there is a limit to how high we can push $g_2$ without perturbation theory breaking down in AdS space.  Even without Planck brane gauge kinetic terms, the largest $g_2$ can be is $4\pi /\sqrt{\log \epsilon^{-1}}$, assuming $g_\rho \lsim 4\pi$.   If $\epsilon \sim 10^{-15}$, then $g'$ cannot be much larger than $3.5 g_{EW}$.  Therefore, if we want the $t'$ and $W'$ to be roughly degenerate, we have to shrink the size of AdS space.  In CFT language, the $SU(2)_i$ beta function is very large in a large $N$ CFT:
\be
\frac{b_{CFT}}{8\pi^2} \sim \frac{1}{g_\rho^2} \sim \frac{N}{16\pi^2},
\ee
so all gauge coupling run to zero in the infrared.  If we leave the confinement scale fixed, the way to increase gauge couplings is to have gauge coupling running begin at a lower scale.  We will explore the possibility of shrinking the conformal window more thoroughly in Section \ref{section:Phenomenology}.

Even if we do decrease $1/L_0$, we may still want to increase the separation between the partner masses and the electroweak scale.  In the limit $m_{t'} = m_{q'}$, the fermion contibution to the Higgs quartic coupling is \cite{Katz:2003sn}
\be
\delta_{\rm fermion} \lambda_h = \frac{\lambda_{\rm top}^4}{\pi^2},
\ee
so ignoring the gauge sector, the electroweak scale would not be very different from $f_\pi$:
\be
\label{eqn:naivevew}
v_{EW} = \sqrt{\frac{-m_h^2}{\lambda_h}} = \sqrt{3}f_\pi.
\ee
Of course, there is a positive contribution to $m_h^2$ coming from the gauge sector  \cite{Arkani-Hamed:2002qy}
\be
\label{eqn:Gauge Log Correction}
\delta_{\rm gauge} m_h^2 = \frac{9}{64\pi^2}g_{EW}^2 f^2 \log \frac{m_\rho^2}{m_{W'}^2},
\ee
which grows large if we are able to increase the mass of $W'$, so $v_{EW}$ will certainly be smaller than the value in equation (\ref{eqn:naivevew}).  Also, there are other sources of $SU(5)$ symmetry breaking in a realistic model that would tend to give a positive contribution to $m_h^2$, such as the inclusion of $U(1)_Y$ effects and the (unspecified) mechanism to remove the $\eta$ field from the spectrum.

Taken together, the final $m_h^2$ value will be
\be
m_h^2 = \delta_{\rm fermion} m_h^2 + \delta_{\rm gauge} m_h^2 + \delta_{\rm other} m_h^2.
\ee
The first two pieces are finite and calculable in our model, and while in principle $\delta_{\rm other} m_h^2$ could be calculable if the additional $SU(5)$ violating effects are nonlocal in AdS, in general we expect the IR brane to be a complicated place which could admit small sources of local $SU(5)$ violation which might not be unambiguously determined.  In order to increase the separation between $v_{EW}$ and $f_\pi$, we will present calculations where we allow ourselves to add $\delta_{\rm other} m_h^2$ contributions, but we will limit ourselves to $10\%$ fine-tuning, meaning that $|m_h^2| / |\delta m_h^2|$ must be greater than $.1$ for each individual contribution.  We will also show that in certain regions of parameter space, $\delta_{\rm fermion} m_h^2$ and $\delta_{\rm gauge} m_h^2$ can balance each other without including a $\delta_{\rm other} m_h^2$ piece.

We remark that this philosophy towards tuning is very different than the one presented in  \cite{Agashe:2004rs}.  In our case, we have a natural mechanism for generating a large Higgs quartic coupling, so for fixed electroweak scale, the Higgs mass $m_{h_0} = \sqrt{2\lambda_h} v_{EW}$ is reasonably heavy.  However, because $\lambda_{\rm top}$ is large, we have to do some amount of tuning to increase the separation between $v_{EW}$ and $f_\pi$.  In \cite{Agashe:2004rs}, some amount of tuning between fermion contributions to the radiative potential is needed to get a hierarchy between $v_{EW}$ and $f_\pi$, and this tuning does not yield a very large Higgs quartic, so the Higgs boson is correspondingly very light (though within experimental bounds).  In our model, we require additional sources of symmetry breaking to avoid precision electroweak constraints, and in their model, they require additional sources of symmetry breaking to allow for a greater range in the Higgs mass.  However, whereas we posit additional unknown sources of symmetry breaking to decrease the absolute value of a relevant parameter ($-m_h^2$), they introduce additional interactions to increase the value of a marginal parameter ($\lambda_h$).   In addition, we expect that a similar level of tuning would be necessary in \cite{Agashe:2004rs} even if $\lambda_{\rm top}$ and $g_{EW}$ were smaller, whereas in our case, the ratio  between $v_{EW}$ and $f_\pi$ is naturally on the order of these small parameters.

\section{Collective Breaking in AdS Space}
\label{section:collective}

The Higgs doublet receives a mass and quartic coupling via quantum corrections. To evaluate these corrections, we will calculate the Coleman-Weinberg potential \cite{Coleman:jx} to one-loop order in a background with non-zero $\Sigma$.    Expanding the potential $V(\Sigma)$ in the Goldstone fields, we can easily identify the quantum corrections to the Higgs mass and quartic coupling.  At the end of the day, we will take the $f_4 \rightarrow \infty$ limit in order to decouple all information about how $SU(5)$ is broken to $SO(5)$ on the IR brane.  In the next section, we will present numerical calculations with specific values of the parameters in order to illustrate the tensions mentioned in the Section \ref{sec:tensions}.  In this section, we focus on trying to understand the structure of the Coleman-Weinberg integrals.

What is most fascinating about the littlest Higgs in AdS space is that the expressions for the mass and quartic coupling manifestly exhibit collective breaking. The 5D Coleman-Weinberg potential is a function of 5D propagators evaluated on the IR brane (See Appendix \ref{conventions} for the form of these propagators).  In both the gauge boson and fermion case, there are two relevant (rescaled) propagators which we can write schematically as
\be
G = \hat{G}^{(+,+)} (p; L_1,L_1), \qquad G_{\mathrm{br}} = \hat{G}^{(-,+)} (p; L_1,L_1),
\ee
corresponding to fields with $(+,+)$ and $(-,+)$ boundary conditions.  If all of the fields that coupled to $\Sigma$ had the same boundary conditions on the UV brane, then $SU(5)$ would be a good symmetry everywhere in AdS space and the Higgs would be an exact Goldstone.  The degree to which $SU(5)$ is broken on the UV brane tells us the degree to which the Higgs is a pseudo-Goldstone boson, so the expression for the Higgs mass and quartic coupling will be proportional to $(G_\mathrm{br}-G)$.  Indeed, by collective breaking, the expressions will be proportional to \emph{two} factors of $(G_\mathrm{br}-G)$, showing that two ``coupling constants'' have to be non-zero for the Higgs to acquire a radiative potential.

We begin with the gauge sector.  The gauge bosons couples to $\Sigma$ on the IR brane through the $\Sigma$ kinetic term:
\be
\label{eqn:sigmagauge}
\mathcal{L}_{\Sigma} = \sqrt{-g_{\mathrm{ind}}} \delta(z-L_1) g_{\mathrm{ind}}^{\mu \nu} \frac{f_5^2}{8} \tr(D_{\mu} \Sigma)^\dagger(D_{\nu} \Sigma)  \supset \\ \epsilon^2 \frac{f_5^2 g_5^2 }{8} A^{\mu a}A^b_{\mu} \tr(T_a \Sigma + \Sigma T_a^T)(T_b^T \Sigma^\dagger + \Sigma^\dagger T_b).
\ee
The (rescaled) mass-squared matrix for the gauge boson in the background $\Sigma$ is therefore
\begin{equation}
 M^2_{ab} = \frac{f_5^2 g_5^2}{4} \tr(T_a \Sigma + \Sigma T_a^T)(T_b^T \Sigma^\dagger + \Sigma^\dagger T_b).
\end{equation}
To generate the effective potential for $\Sigma$ to 1-loop order, we consider an arbitrary number of mass insertions, analytically continuing the propagators to Euclidean space:
\be
\label{eqn:cwgauge}
\delta_{\rm gauge} \left(\sqrt{-g_{\mathrm{ind}}} V(\Sigma)\right)= \frac{3}{2} \int \frac{ p^3 dp}{8 \pi ^2} \tr \sum_{n=0}^{\infty} \frac{(-1)^n}{n}(\hat{G} \cdot M^2)^n =  \frac{3}{2} \tr \int \frac{ p^3 dp}{8 \pi ^2} \log(1 + \hat{G} \cdot M^2),
\ee
where $\hat{G}_{ab} = \hat{G}_a(p;L_1,L_1) \delta_{ab}$ are the rescaled propagators for the $SU(5)$ gauge bosons, and $a$ labels the generators of $SU(5)$.  (The factor of $3$ accounts for the three polarizations of each gauge boson.)  The propagators for the $SU(2)_i$ subgroups will be labeled $G_i$ for $i = 1,2$, and the propagators for the rest of $SU(5)$ will be labeled $G_\mathrm{br}$ to indicate that those propagators have Dirichlet boundary conditions on the UV brane.

Using the fact that $\tr \log X =  \log \det X$, we can evaluate equation (\ref{eqn:cwgauge}) exactly.  Expanding $V(\Sigma)$ in powers of the Goldstone fields, the mass correction to the Higgs potential from gauge bosons loops is (including wavefunction renormalization for the Higgs):
\begin{equation}
\delta_{\rm gauge} m_h^2 = \frac{9}{16 Z_h \epsilon^2} \int \frac{ p^3 dp}{8 \pi ^2} \frac{f_5^2 g_5^4 \left(G_{\mathrm{br}}- G_1\right)\left( G_\mathrm{br}- G_2\right)}{(1+f_5^2 g_5^2 G_\mathrm{br}/2)(1+f_5^2 g_5^2 (G_1+G_2)/4)}.
\end{equation}
In the numerator, we manifestly see collective breaking, in that both $G_1 \neq G_{\rm br}$ and $G_2 \neq G_{\rm br}$ in order for the Higgs to get a radiative potential from the gauge sector.  As already mentioned, we want to take the $f_5 \rightarrow \infty$ limit.  The resulting expression only depends on the AdS parameters through $f_\pi = 2/(L_1 g_\rho)$ and  the propagators themselves:
\begin{equation}
\label{eqn:higgsmass}
\delta_{\rm gauge} m_h^2 =  \frac{9 }{2 f_\pi^2} \int \frac{ p^3 dp}{8 \pi ^2} \frac{\left(G_{\mathrm{br}}- G_1\right)\left( G_\mathrm{br}- G_2\right)}{G_\mathrm{br}(G_1+G_2)}.
\end{equation}
We will discuss the physical meaning of this integrand after we calculate the Higgs quartic.

In the littlest Higgs, there is a radiatively generated trilinear coupling $\phi h h$ and $\phi^2$ mass term.  Looking the at the fermion coupling in equation (\ref{eqn:yukawainteraction}) and the fermion content in equation (\ref{eqn:FermionTable}), it is clear that there is still a global $SU(2)$ symmetry present in the fermion sector that protects the triplet from radiative corrections, so all contributions to the $\phi$ potential will come from gauge boson loops.\footnote{In particular, in this model there is no danger of having a negative triplet mass squared.}  Using the Coleman-Weinberg potential we can calculate the contribution to the triplet mass and $\phi hh$ coupling to 1-loop order in the $f_5 \rightarrow \infty$ limit. To go to this limit, we need to canonically normalize $\phi$, and we use $Z_h = Z_\phi$:
\be
\label{eqn:gaugephiintegrand}
\delta_{\rm gauge} m^2_{\phi} = \frac{3}{f_\pi^2}\int \frac{dp p^3}{8\pi^2} \frac{2G_{\rm br}^2 + 4 G_1 G_2 - 3 G_{\rm br}(G_1 + G_2)}{G_{\rm br}(G_1 + G_2)},
\ee
\be
\label{eqn:gaugeyukawaintegrand}
\delta_{\rm gauge} \lambda_{\phi hh} = \frac{3 }{2 f_\pi^3}\int \frac{dp p^3}{8\pi^2} \frac{G_1-G_2}{G_1+G_2}.
\ee
In the case $G_1 = G_2$, we see that no $\phi h h$ coupling is generated, corresponding to the $T$-parity limit of the theory.  The direct contribution of the gauge bosons to the Higgs quartic is
\begin{equation}
\label{eqn:higgsquartic}
\delta_{\rm gauge} \lambda'_h =  -\frac{3}{4 f_\pi^4} \int \frac{ p^3 dp}{8 \pi ^2} \frac{3 G_{\rm br}^2(G_{\rm br}^2 -G_1^2-G_2^2) + G_{\rm br} (G_{\rm br}^2 - G_1 G_2) (G_1 + G_2) + 3 G_1^2 G_2^2   }{G_{\rm br}^2 (G_1+G_2)^2}.
\end{equation}
After integrating out the heavy triplet, the total gauge contribution to the Higgs quartic is
\be
\delta_{\rm gauge} \lambda_h = \delta_{\rm gauge} \lambda'_h -  \frac{(\delta_{\rm gauge} \lambda_{\phi hh} )^2}{\delta_{\rm gauge} m^2_{\phi}}.
\ee
Collective breaking is certainly not manifest in this form, but we can check that collective breaking occurs slice by slice in momentum space.  Imagine doing each momentum integral from $p_0$ to $p_0 + \Delta p_0$.  The contribution from this slice to $\delta_{\rm gauge} \lambda_h$ is
\be
\left. \frac{1}{f_\pi^4} \frac{\Delta p_0 p_0^3}{8\pi^2} \frac{\left(G_{\mathrm{br}}- G_1\right)\left( G_\mathrm{br}- G_2\right)}{G_{\rm br}^2 (G_1+G_2)^2} \frac{E_4(G_{\rm br}, G_1, G_2)}{E_2(G_{\rm br}, G_1, G_2)} \right|_{p=p_0},
\ee
where $E_i$ are unenlightening $i$-th order polynomials.  We see readily that the gauge contribution to  $\lambda_h$ vanishes in each slice of momentum space unless both $G_1$ and $G_2$ are different from $G_{\rm br}$.

%\begin{figure}
%\begin{center}
%\includegraphics[scale=.85]{gaugehiggsmass}
%\end{center}
%\caption{The gauge boson integrand for $\delta_{\rm gauge}m_h^2$ using the parameters of Model 1c.  The linear behavior near $p=0$ corresponds to the quadratic sensitivity of this operator below the mass of the $W'$. The $W'$ has mass around $.18 /L_1$ and the $\rho$-like resonances appear at $2.4/L_1$.}
%\label{figure:gaugehiggsmass}
%\end{figure}

\FIGURE[t]{
\parbox{5in}{\centering
\includegraphics[scale=.85]{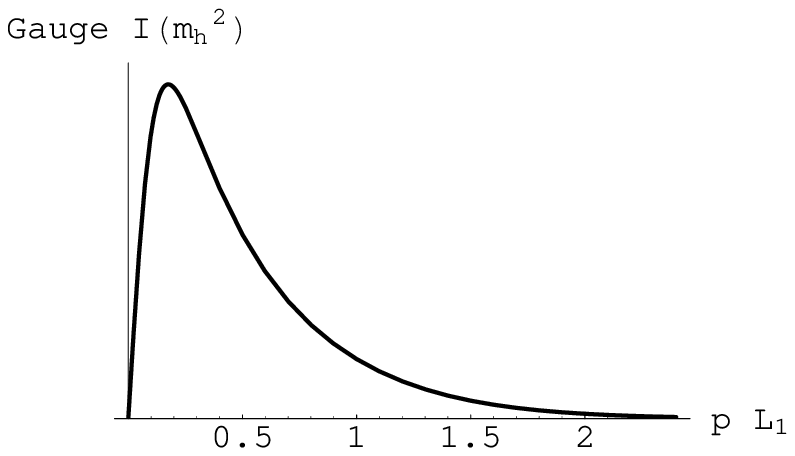}
}
\caption{The gauge boson integrand for $\delta_{\rm gauge}m_h^2$ using the parameters of Model 1c.  The linear behavior near $p=0$ corresponds to the quadratic sensitivity of this operator below the mass of the $W'$. The $W'$ has mass around $.18 /L_1$ and the $\rho$-like resonances appear at $2.4/L_1$.}
\label{figure:gaugehiggsmass}
}

Before proceeding to analyze the fermion contribution, it is interesting to ask whether these 1-loop contributions match our expectations from the low energy theory.  Expanding the integrand in equation (\ref{eqn:higgsmass}) to lowest order in momenta and integrating with a hard momentum cutoff $\Lambda_0$:
\be
\label{eqn:lowenhiggsmass}
\delta_{\rm gauge} m_h^2 \sim \int^{\Lambda_0}\! I(p\rightarrow 0)\, dp  =  \frac{9g_{EW}^2 \Lambda_0^2}{64\pi^2} + \mathcal{O}(\Lambda_0^4/f_\pi^2),
\ee
which is exactly the expected quadratically divergent contribution to the Higgs mass from a $W$ boson loop.   Figure \ref{figure:gaugehiggsmass} is a plot of the integrand of equation (\ref{eqn:higgsmass}) with the parameters of Model 1c from the next section. At low energies, we see the linear behavior in $p$ corresponding to the quadratic divergence from the $W$ loops. At momenta around $m_{W'} \sim .18 / L_1$, this divergence is softened by the appearance of the $W'$ partner.  The $1/p$ behavior between $m_{W'}$ and $m_\rho \sim 2.4 /L_1$ reflects a logarithmic threshold correction, and at $p \sim m_\rho$, the integrand dies off exponentially fast.

Note that the peak in the $m_h^2$ integrand is almost exactly at $m_{W'}$.  More generally, when there is a quadratic divergence at low energies that is cutoff by the existence of new states, the integrands are roughly of the form (ignoring coupling constants):
\be
I(p) \sim \frac{p^3}{8\pi^2} \frac{1}{ f_\pi^2 p^2} \prod_{i} \left(1 -  \frac{p^2}{p^2+ m_i^2}  \right) \sim \frac{p^3}{8\pi^2} \frac{1}{f_\pi^2 p^2} \left(1 -  \frac{p^2}{p^2+ m_{\rm partner}^2}  \right)^N,
\ee
where $m_i$ is the mass of any partner particle whose appearance would cutoff the quadratic divergence, and in the last step, we have gone to the limit where all $N$ of the partners are degenerate with mass $m_{\rm partner}$.  The peak in $I(p)$ is at
\be
p \sim \frac{m_{\rm partner}}{\sqrt{2N-1}}.
\ee
For the gauge quadratic divergence to the Higgs mass, $N = 1$ because there is only the $W'$ partner particle, and indeed the peak in Figure \ref{figure:gaugehiggsmass} is at the mass of the $W'$.  We can do the integral over $p$ of $I(p)$:
\be
\int \! I(p) \, dp= \frac{m^2_{\rm partner}}{(4\pi f_\pi)^2}\log \frac{\Lambda^2}{m_{\rm partner}^2} \quad(N=1), \qquad \frac{m^2_{\rm partner}}{(4\pi f_\pi)^2} \frac{1}{N-1} \quad(N>1).
\ee
These factors are what a low energy observer would call ``unknown $\mathcal{O}(1)$ coefficients'' that multiply quadratically sensitive operators.  In the example of a single $W'$ ($N=1$), there is still a logarithmic divergence, and indeed the $\rho$-like resonances provide an effective cutoff $\Lambda \sim m_\rho$ for $\delta_{\rm gauge} m_h^2$.

%\begin{figure}
%\begin{center}
%\includegraphics[scale=.85]{gaugephimass} $\qquad$ \includegraphics[scale=.85]{gaugephiyukawa}
%\end{center}
%\begin{center}
%\end{center}
%\begin{center}
%\includegraphics[scale=.85]{gaugehiggsquartic}
%\end{center}
%\caption{The gauge boson integrands for $\delta_{\rm gauge}m_\phi^2$, $\delta_{\rm gauge}\lambda_{\phi h h }$, and $\delta_{\rm gauge}\lambda_{h}'$ using the parameters of Model 1c.  The mass of the first $\rho$-like resonances is $\sim 2.4/L_1$, and all of these integrands become exponentially small at momenta corresponding to the second $\rho$-like resonances $\sim 5.5/L_1$.  See the text for why the peaks of these integrands are at $p \sim 1.3/L_1$.  Note the $-1/p$ behavior of $\delta_{\rm gauge}\lambda_{h}'$ near $p=0$, corresponding to expected low energy logarithmic running of the Higgs quartic.}
%\label{figure:gaugeother}
%\end{figure}

\FIGURE[t]{
\parbox{6.0in}{\centering
\includegraphics[scale=.85]{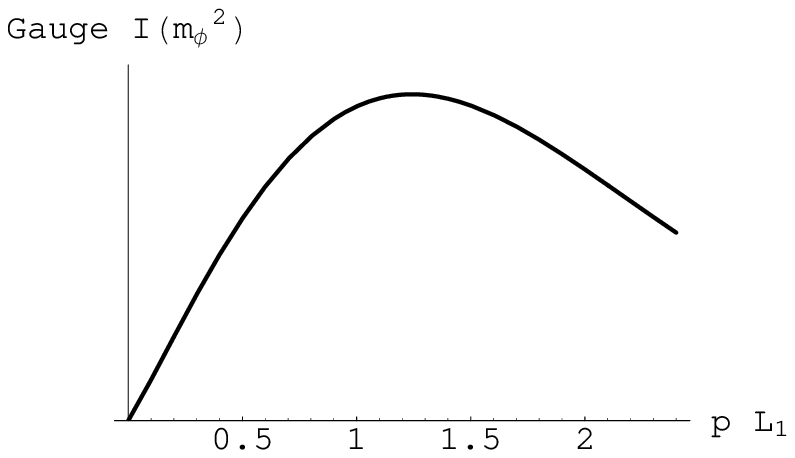} $\qquad$ \includegraphics[scale=.85]{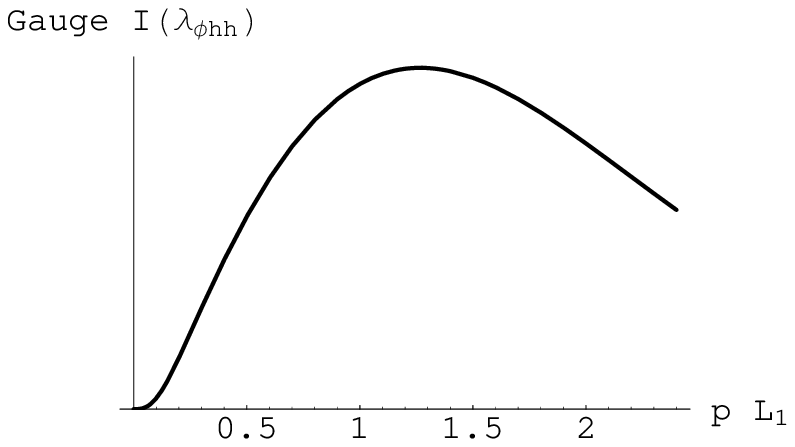}}

\parbox{6.0in}{\centering
\includegraphics[scale=.85]{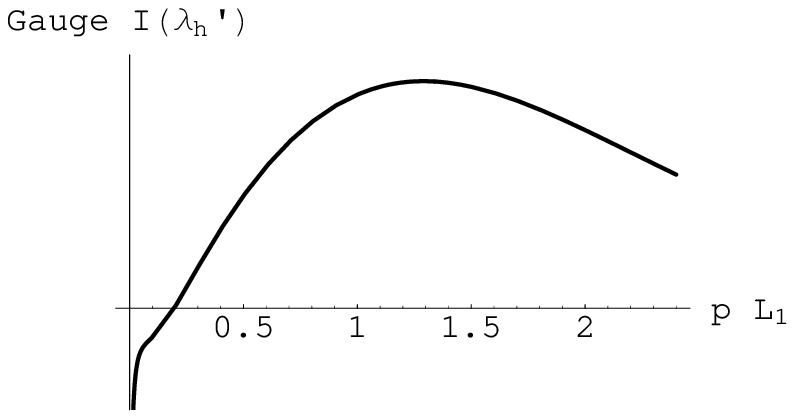}
}
\caption{The gauge boson integrands for $\delta_{\rm gauge}m_\phi^2$, $\delta_{\rm gauge}\lambda_{\phi h h }$, and $\delta_{\rm gauge}\lambda_{h}'$ using the parameters of Model 1c.  The mass of the first $\rho$-like resonances is $\sim 2.4/L_1$, and all of these integrands become exponentially small at momenta corresponding to the second $\rho$-like resonances $\sim 5.5/L_1$.  See the text for why the peaks of these integrands are at $p \sim 1.3/L_1$.  Note the $-1/p$ behavior of $\delta_{\rm gauge}\lambda_{h}'$ near $p=0$, corresponding to expected low energy logarithmic running of the Higgs quartic.}
\label{figure:gaugeother}
}

The expressions for the quartic coupling are also very interesting.  Expanding equations (\ref{eqn:gaugephiintegrand}), (\ref{eqn:gaugeyukawaintegrand}) and (\ref{eqn:higgsquartic}) in momenta and integrating up to $\Lambda_0$:
\be
\nonumber
\delta_{\rm gauge} m^2_\phi \sim \frac{3 g_{EW}^2 }{8\pi^2}\Lambda_0^2 + \mathcal{O}(\Lambda_0^4/f_\pi^2),\qquad
 \delta_{\rm gauge} \lambda_{\phi h h} \sim \frac{3 (g_1^2 - g_2^2)}{64 \pi^2 (g_1^2 + g_2^2) } \frac{ \Lambda_0^4}{f_\pi^3} + \mathcal{O}(\Lambda_0^6/f_\pi^5),
\ee
\be
\label{gaugequarticlowenergyint}
 \delta_{\rm gauge} \lambda'_{h} \sim -\frac{9 g_{EW}^4}{256 \pi^2} \log \Lambda_0^2 + \mathcal{O}(\Lambda_0^2/f_\pi^2).
\ee
To leading order in $\Lambda_0$, $\delta_{\rm gauge} \lambda_h$ comes entirely from $\delta_{\rm gauge} \lambda'_{h}$.  This logarithmic divergence is the standard contribution to a doublet scalar from a $W$ boson loop.  Looking at the integrands in Figure \ref{figure:gaugeother}, we see that each one rises linearly until it peaks at $\sim 1.3/L_1$.  The linear behavior reflects the fact that the low energy operators which generate the Higgs quartic coupling are quadratically sensitive to the cutoff.  However, there is no state in the spectrum at $\sim 1.3/L_1$ where the quadratic divergence is softened.  Rather, the $\rho$-like states appear at $\sim 2.4/L_1$ but $N=2$ because along with the $(+,+)$ gauge bosons, the appearance of the first KK mode of the $(-,+)$ gauge bosons or the second KK mode of the $(+,-)$ gauge bosons would restore enough of the $SU(5)$ global symmetry to protect the Higgs mass.\footnote{The $(+,-)$ modes technically only exist in the $f_4 \rightarrow \infty$ limit.  The first KK $(+,-)$ mode is the $W'$ and the second KK mode is another $\rho$-like state which is roughly degenerate with the first $(-,+)$  KK mode.  We can either restore $SU(5)$ near the UV brane or near the IR brane, and in this sense $N=2$ and $m_{\rm partner} = m_\rho \sim g_\rho f_\pi$.}    At momenta $p \sim 5.5/L_1$ the integrands are exponentially suppressed, as one might expect as this is the scale of the next-to-lightest KK gauge boson modes.

Now for the fermion contribution to the Higgs potential. The bulk fermions couple to the IR brane scalar $\Sigma$ via
\begin{equation}
 \mathcal{L}_{\mathrm{Yukawa}} = \sqrt{-g_{\mathrm{ind}}} \delta(z-L_1)(\lambda \bar{\mathbf{Q}} \Sigma \mathbf{Q^c} + h.c.).
\end{equation}
Unlike equation (\ref{eqn:yukawainteraction}), we are explicitly using bulk Dirac notation, though recall that the boundary conditions eliminate the lower (upper) Weyl fermion in $\mathbf{Q}$ ($\mathbf{Q^c}$) on the IR brane.   In order to form a fermion loop and contract the fields, we need two such Yukawa insertions, so the one-loop effective potential for $\Sigma$ is
\be
\label{eqn:cwfermion}
\delta_{\rm fermion} \left( \sqrt{-g_{\mathrm{ind}}}V(\Sigma) \right) =  -3 \tr \int \frac{ p^3 dp}{8 \pi ^2} \log(1 + \hat{S} \cdot M \cdot \hat{S}^c \cdot M^\dagger),
\ee
where $\hat{S}_{ab} = \hat{S}_a \delta_{ab}$ are bulk 5D propagators (with Dirac indices) for $\mathbf{Q}$, and similarly for $\hat{S}^c$.  The index $a$ runs over a fundamental of $SU(5)$, and the rescaled mass matrix in a background $\Sigma$ is
\begin{equation}
M_{ab} = \frac{\lambda}{\epsilon} \Sigma_{ab}.
\end{equation}
The trace in equation (\ref{eqn:cwfermion}) runs over both $SU(5)$ and Dirac indices, and the factor of 3 takes into account the $SU(3)_C$ charges of $\mathbf{Q}$ and $\mathbf{Q^c}$.

Because of our choice of boundary conditions on the IR brane, we can easily do the trace over Dirac indices.  From Appendix \ref{conventions}, the (rescaled) bulk fermion propagators take the schematic form
\begin{equation}
\hat{S}(p;L_1,L_1;\nu) = (A \slashed{p} + B \gamma^5 + C ) P_L, \qquad \hat{S}^c(p;L_1,L_1;\nu^c) = (A^c \slashed{p} + B^c \gamma^5 + C^c ) P_R,
\end{equation}
where $P_{L,R} = (1\mp \gamma^5)/2$.  Using the trace properties of the $\sigma$ matrices:
\begin{equation}
 \hat{S} = \left(%
\begin{array}{cc}
  C - B & 0 \\
A \bar{\sigma} \cdot p & 0 \\\end{array}%
\right), \qquad
\hat{S^c} = \left(%
\begin{array}{cc}
  0 & A^c \sigma \cdot p \\
0 & C^c+B^c \\\end{array}%
\right), \qquad \Tr (\hat{S} \hat{S}^c)^n = 2 (p^2 A A^c)^n.
\end{equation}
Note that $A = \hat{G}(p; L_1,L_1; \nu)$, so equation (\ref{eqn:cwfermion}) can be rewritten as
\be
\delta_{\rm fermion} \left( \sqrt{-g_{\mathrm{ind}}}V(\Sigma) \right) =  -3 \tr \int \frac{ p^3 dp}{8 \pi ^2} 2 \log(1 + p^2 \hat{G} \cdot M \cdot \hat{G}^c \cdot M^\dagger),
\ee
where now the trace runs only over $SU(5)$ indices.

We can proceed as in the gauge boson case and compute the Higgs mass and quartic coupling.  We designate propagators for modes with $(+,+)$ boundary conditions as $G$ and $G^c$, and with $(-,+)$ boundary conditions as $G_{\rm br}$ and $G^c_{\rm br}$.  These propagators are functions of $\nu$ and $\nu^c$, respectively.  After canonically normalizing $h$ and taking the $f_4 \rightarrow \infty$ limit:
\begin{equation}
\label{eqn:higgsmassfer}
\delta_{\rm fermion} m_h^2 = -\frac{12}{f_\pi^2} \int \frac{ p^3 dp}{8 \pi ^2} \frac{\hat{\lambda}^2 p^2(G_\mathrm{br}- G)(G^c_\mathrm{br} - G^c)} {(1 + \hat{\lambda}^2 p^2 G G^c_\mathrm{br})(1 + \hat{\lambda}^2 p^2 G_\mathrm{br}G^c)},
\end{equation}
\begin{equation}
\label{eqn:higgsquarticfer}
\delta_{\rm fermion} \lambda _h = -\frac{4 \lambda^2}{f^4_\pi}\int \frac{ p^3 dp}{8 \pi ^2} \frac{ p^2(G_\mathrm{br}- G)(G^c_\mathrm{br} - G^c) }{\left( 1 +  p^2 \hat{\lambda}^2 G G^c_\mathrm{br} \right)^2 \left( 1 +  p^2 \hat{\lambda}^2 G_\mathrm{br} G^c \right)^2} F(G, G_\mathrm{br}, G^c, G^c_\mathrm{br}),
\end{equation}
where $\hat{\lambda} = \lambda/ \epsilon$ and $F(G, G_\mathrm{br}, G^c, G^c_\mathrm{br})$ is
\begin{equation}
F(G, G_\mathrm{br}, G^c, G^c_\mathrm{br}) = 4+ \hat{\lambda}^2 p^2(3GG^c + 3 G_\mathrm{br}G^c_\mathrm{br} + G G^c_\mathrm{br}+ G_\mathrm{br} G^c ) + 4 \hat{\lambda}^4 p^4G G_\mathrm{br} G^c G^c_\mathrm{br}.
\end{equation}
These expressions manifestly exhibit collective breaking.

\FIGURE[t]{
\parbox{6.0in}{\centering
\includegraphics[scale=.85]{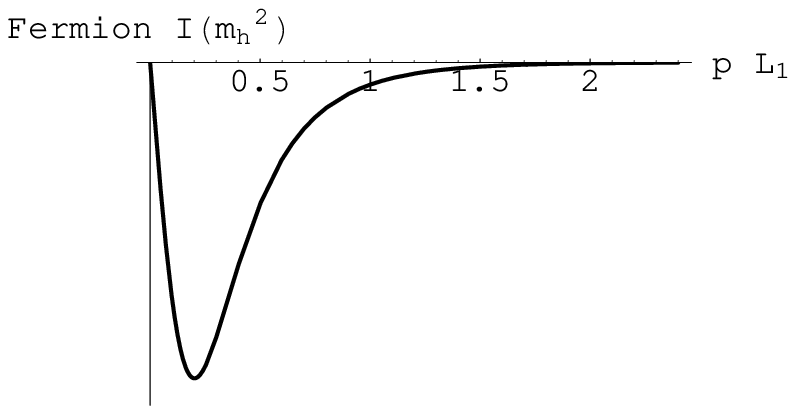} $\qquad$
\includegraphics[scale=.85]{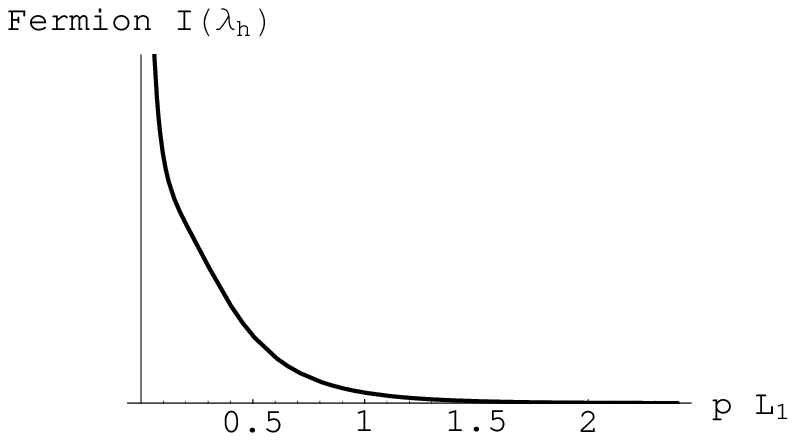}
}
\caption{The fermion integrands for $\delta_{\rm gauge}m_h^2$ and $\delta_{\rm gauge}\lambda_h$ using the parameters of Model 1c.  The linear behavior of the mass integrand and the $1/p$ behavior of the quartic integrand reflect the expected quadratic and logarithmic sensitivity of these operators at energies beneath the mass of the fermion partners.  The $q'$ and $t'$ are degenerate with mass $\sim .36/L_1$.  See the text for why the peak of the mass integrand is at $p \sim .2/L_1$.}
\label{figure: fermionhiggs}
}

%\begin{figure}
%\begin{center}
%\includegraphics[scale=.85]{fermionhiggsmass} $\qquad$
%\includegraphics[scale=.85]{fermionhiggsquartic}
%\end{center}
%\caption{The fermion integrands for $\delta_{\rm gauge}m_h^2$ and $\delta_{\rm gauge}\lambda_h$ using the parameters of Model 1c.  The linear behavior of the mass integrand and the $1/p$ behavior of the quartic integrand reflect the expected quadratic and logarithmic sensitivity of these operators at energies beneath the mass of the fermion partners.  The $q'$ and $t'$ are degenerate with mass $\sim .36/L_1$.  See the text for why the peak of the mass integrand is at $p \sim .2/L_1$.}
%\label{figure: fermionhiggs}
%\end{figure}

It is again very instructive to match the results with our expectations from effective field theory. As in the bosonic case, we can expand the integrands to lowest order in momenta and integrate with a hard momentum cutoff $\Lambda_0$:
\be
\delta_{\rm fermion}m_h^2 \sim -\frac{3 \lambda_{\rm top}^2 }{8 \pi^2}\Lambda_0^2 + \mathcal{O}(\Lambda_0^4/f_\pi^2), \qquad
\delta_{\rm fermion} \lambda _h \sim \frac{3 \lambda_{\rm top}^2}{16 \pi^2} \log \Lambda_0^2+ \mathcal{O}(\Lambda_0^4/f_\pi^4).
\ee
These are precisely the low energy expectations for the contribution of top loops to the Higgs doublet mass and quartic.  The integrands from equations (\ref{eqn:higgsmassfer}) and (\ref{eqn:higgsquarticfer}) appear in Figure \ref{figure: fermionhiggs}.  The mass integrand starts off  linearly, corresponding to the low energy quadratic divergence.  In Model 1c, the $q'$ and $t'$ are degenerate with mass $\sim .36/L_1$, so $N=2$ and the peak in the integrand is at $p \sim .2/L_1$ as expected.  The quartic integrand is dominated by the $1/p$ piece corresponding to the logarithmic low energy behavior, though there is a slight bump at the mass of the $q'$ and $t'$.  Unlike the quartic integrands in the gauge sector, these integrals are already very suppressed at the mass of the KK fermion modes $\sim \pi/L_1$, which is as expected because the quartic contribution is finite at one-loop with the fermion content of \cite{Katz:2003sn}.

\section{Numerical Examples}
\label{section:Phenomenology}

%\begin{figure}
%\begin{center}
%\includegraphics[scale=.85]{higgsmassone} $\qquad$
%\includegraphics[scale=.85]{higgsmasstwo}
%\end{center}
%\caption{Values of the physical Higgs mass in Model 1, keeping $f_\pi = 1 \TeV$ fixed and varying the bulk masses $\nu$ and $\nu^c$ ($g' = 2g_{EW}$).  Outside of the region $|\nu|, |\nu^c| \lsim .4,$ it is difficult to satisfy the condition $\lambda_{\rm top} = 1$.}
%\label{figure:allvaluesnu}
%\end{figure}

\FIGURE[t]{
\parbox{6.0in}{
\includegraphics[scale=.85]{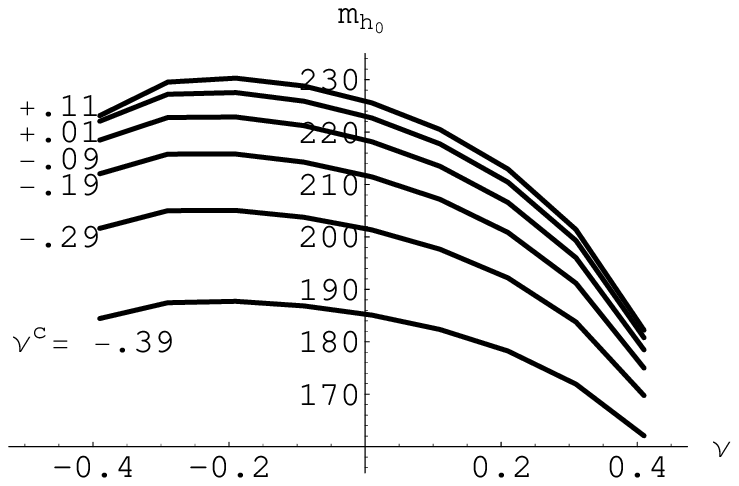} $\qquad$
\includegraphics[scale=.85]{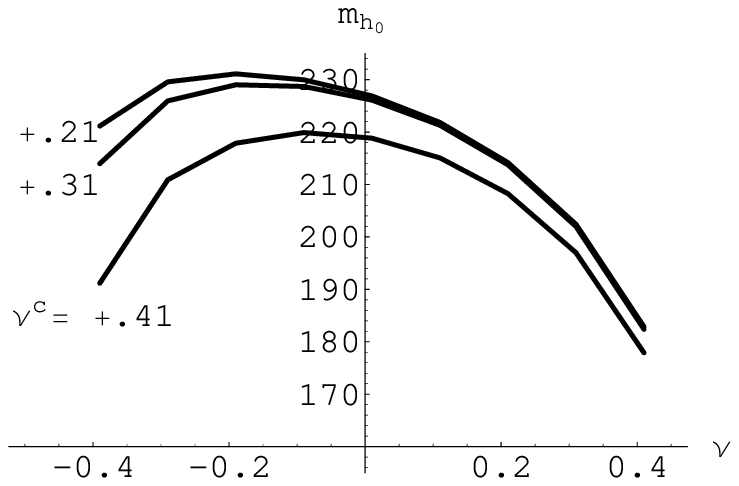}
}
\caption{Values of the physical Higgs mass in Model 1, keeping $f_\pi = 1 \TeV$ fixed and varying the bulk masses $\nu$ and $\nu^c$ ($g' = 2g_{EW}$).  Outside of the region $|\nu|, |\nu^c| \lsim .4,$ it is difficult to satisfy the condition $\lambda_{\rm top} = 1$.}
\label{figure:allvaluesnu}
}

Now that we understand the structure of the Coleman-Weinberg integrals, we want to input specific values of the AdS parameters to find the generic value of the physical Higgs mass in this setup.  In order to deal with the infrared logarithmic divergences in the quartic coupling, we do each Coleman-Weinberg integral from $p = m_{h_0}$ to $\infty$, iterating the calculations until we get a stable $m_{h_0}$ value.  We choose the parameters of our theory to fit the known top Yukawa coupling, electroweak gauge coupling, and electroweak scale.  In Model 1, we allow for an unknown $\delta_{\rm other} m_h^2$ contribution to the Higgs potential in order to fix the pion decay constant at $f_{\pi} = 1 \TeV$, and this will generically require fine-tuning at the 10\% level.  In Model 2, we arrange the gauge and fermion contributions to the doublet mass to cancel against each other with no considerable fine-tuning by shrinking the conformal window, but in these models $f_\pi$ will be significantly lower, and hence the $t'$, $q'$ and $W'$ partners will be generically lighter.  Finally in Model 3, we show a compromise where we shrink the conformal window but still allow for a $\delta_{\rm other} m_h^2$ piece, yielding sufficiently heavy partner particles with no considerable fine-tuning (25\%).  

We begin in Model 1 by choosing the brane separation to generate the full hierarchy between $M_\Pl$ and $f_\pi$, namely $\epsilon \sim 10^{-15}$. Once this AdS geometry is fixed, it is clear from equation (\ref{eqn:LowEnergyGaugeCoupling}) that in order for $g_{\rho}$ to remain perturbative, we need the gauge kinetic terms on the UV boundary to be close to zero (\emph{i.e.}\ choose the Landau pole of the gauge couplings to be as close to the UV as possible) and therefore we must take approximately $g_1 = g_2$.   In terms of the electroweak gauge coupling, $g' = 2g_{EW}$.   When we do consider cases where $g_1 \not= g_2$, we take $z_2 = 0$ for simplicity.  We still have the freedom to choose the bulk masses of the fermions. Looking at equation (\ref{eqn:hubris}), the requirement of $\lambda_{\rm top}=1$ restricts the range of $\nu$ and $\nu^c$ to around $|\nu|,|\nu^c| \lsim.4$.  As remarked earlier in equation (\ref{eqn:Fermion Log correction}), to minimize the logarithmic correction to $m_h^2$ from the fermion sector, we want the $t'$ and $q'$ fermion partners to be roughly degenerate, and this corresponds to the limit $\nu =-\nu^c$.  To make the $t'$ and $q'$ as light as possible, we want to send $\nu$ as large as possible while still maintaining the desired top Yukawa coupling, corresponding to $\nu\rightarrow 0.4$.

\TABLE[t]{
\begin{tabular}{c|c|c|c}
 & Model 1a & Model 1b & Model 1c \\
\hline
$L_1$ & $(2.6 \TeV)^{-1} $& $(2.6 \TeV)^{-1} $ & $(5.2 \TeV)^{-1} $  \\
$\epsilon$ & $10^{-15}$ & $10^{-15}$ & $10^{-15}$ \\
$g_{EW}$ & 0.63&0.63 &0.63 \\
$v_{EW}$ & $246 \GeV$ & $246 \GeV$ & $246 \GeV$ \\
\hline
$g'$ & $2 g_{EW}$ & $2 g_{EW}$ & $3 g_{EW}$ \\
$g_\rho$ & 5.2 & 5.2 & 10.4 \\
$f_\pi$ & $1.0 \TeV$& $1.0 \TeV$ & $1.0 \TeV$\\
$m_\rho$ & $6.2 \TeV$& $6.2 \TeV$ & $12.2 \TeV$\\
\hline
$\lambda_{\rm top}$ & $1.0$  & $1.0$ & $1.0$\\
$\nu $& $+0.42$ & $+0.20$ & $+0.49$  \\
$\nu^c$ & $-0.42$ & $-0.10$ & $-0.49$ \\
$\lambda$ & $0.26$  & $0.23$ & $0.11$\\
\hline
$m_{W'}$ & $630 \GeV$  & $630 \GeV$ & $950 \GeV$  \\
$m_{t'}$ & $2.0 \TeV$ & $3.0 \TeV$& $1.9 \TeV$ \\
$m_{q'}$ & $2.0 \TeV$ & $3.5 \TeV$& $1.9 \TeV$ \\
\hline
$m_{h_0}$ & $160 \GeV$ & $210 \GeV$ & $210 \GeV$ \\
$\delta_{\rm gauge} m_h^2$ & $+(80 \GeV)^2$  & $+(80 \GeV)^2$ & $+(130 \GeV)^2$\\
$\delta_{\rm fermion} m^2_h$ & $-(310 \GeV)^2 $ & $-(440 \GeV)^2 $ &$-(330 \GeV)^2 $ \\
$\delta_{\rm other} m_h^2$ & $+(270 \GeV)^2$  & $+(410 \GeV)^2$ & $+(270 \GeV)^2$\\
$\delta_{\rm gauge} \lambda_h$ & 0.04 & 0.04 & 0.15  \\
$\delta_{\rm fermion} \lambda_h$ & 0.18 & 0.32 & 0.20  \\
\hline
$m_{\phi}$ & $430 \GeV$ & $430 \GeV$ & $1240 \GeV$\\
$\lambda_{\phi hh}$ & --- & --- & $560 \GeV$
\end{tabular}
\caption{Parameters and results for Model 1.  In this model, we allow the addition of a $\delta_{\rm other} m_h^2$ contribution to the Higgs potential in order to set the pion decay constant $f_\pi = 1 \TeV$.  By fixing $\epsilon = 10^{-15}$, we are limited in how high we can push the $g'$ coupling, and hence the mass of the $W'$.  Fine-tuning in this model is usually of the order of 10\% -- 20\%.}
\label{table:model1}
}

The details of Model 1 are presented in Table \ref{table:model1}, with some variations of the parameters to illustrate the stability of these results.  In particular, we allow $g'$ to be larger than $2 g_{EW}$  in Model 1c to raise the $W'$ mass.  With the addition of a $\delta_{\rm other} m_h^2$ contribution to the Higgs potential, there is no problem getting a viable phenomenology which reproduces the electroweak scale while fixing $f_\pi = 1 \TeV$.    Figure \ref{figure:allvaluesnu} is a plot of the physical Higgs mass as a function of $\nu$ and $\nu^c$ when $g' = 2 g_{EW}$.  Generically the physical Higgs falls within the range $m_{h_0} \sim 150-250 \GeV$.   As expected, the lightest physical Higgs occurs when $\nu \sim -\nu^c \sim 0.4$.  Because $\lambda_{\rm top} \sim 1$, the fermionic contribution to the Higgs doublet mass is generically large, ranging from $\sim 300-500\GeV$, whereas for $g' = 2g_{EW}$, the gauge contribution is fixed at $\sim 80 \GeV$.   This large difference between the gauge and fermion contributions to the doublet mass is expected from equations (\ref{eqn:Fermion Log correction}) and (\ref{eqn:Gauge Log Correction}).   Fine-tuning of the mass parameters is at the $\sim 10\%-20\%$ level, and $\delta_{\rm other} m_h^2$ is responsible for canceling nearly all of the fermionic contribution.  The large quartic coupling comes almost entirely from the fermion sector, though the contributions equalize as $g'$ increases.

\TABLE[t]{
\begin{tabular}{c|c|c|c}
 & Model 2a & Model 2b & Model 2c \\
\hline
$L_1$ & $(530 \GeV)^{-1}$  & $(1.7 \TeV)^{-1}$ & $(1.3 \TeV)^{-1}$ \\
$\epsilon$ & $10^{-5}$ & $10^{-5}$ & $10^{-1}$ \\
$g_{EW}$ & 0.63 &0.63 & 0.63 \\
$v_{EW}$ & $246 \GeV$ & $246 \GeV$ & $246 \GeV$ \\
\hline
$g'$ & $2g_{EW}$ & $5g_{EW}$ & $8 g_{EW}$ \\
$g_\rho$ & 3.0 & 10.5 & 7.6 \\
$f_{\pi}$ &  $580 \GeV$ & $420 \GeV$ &  $340 \GeV$ \\
$m_\rho$ &  $1.2 \TeV$ & $5.2 \TeV$ &  $3.0\TeV$ \\
\hline
$\lambda_{\rm top}$ & 1.0 & 1.0 & 1.0 \\
$\nu $& $+0.17$ & $+0.49$ & $+0.49$ \\
$\nu^c$ & $-0.17$ & $-0.49$ & $-0.49$ \\
$\lambda$ & 0.63 & 0.08 &  0.10 \\
\hline
$m_{W'}$ & $360 \GeV$ & $650 \GeV$ & $850 \GeV$ \\
$m_{t'}$ & $820 \GeV$ & $940 \GeV$ & $520 \GeV$ \\
$m_{q'}$ & $820 \GeV$ & $940 \GeV$ & $520 \GeV$ \\
\hline
$m_{h_0}$ &  $120 \GeV$ & $220 \GeV$ &  $260 \GeV$\\
$\delta_{\rm gauge} m^2_h$ &  $+(20 \GeV)^2$ & $+(80 \GeV)^2$ & $+(70 \GeV)^2$ \\
$\delta_{\rm fermion} m^2_h$ & $-(90 \GeV)^2$ & $-(180 \GeV)^2$ &  $-(200 \GeV)^2$\\
$\delta_{\rm gauge}\lambda_h$ & 0.01 & 0.15 & 0.07 \\
$\delta_{\rm fermion}\lambda_h$ & 0.12 & 0.26 & 0.51 \\
\hline
$m_{\phi}$ & $90 \GeV$ & $860 \GeV$ & $880 \GeV$\\
$\lambda_{\phi hh}$ & --- & $800 \GeV$ & $1.1 \TeV$\\
\end{tabular}
\caption{Parameters and results for Model 2.  In this model, we do not set $f_\pi$ by hand, and rely on a natural cancellation between the gauge and fermion sectors to generate the hierarchy between $f_\pi$ and $v_{EW}$.   As such, $f_\pi$ is quite low, and we need to raise $g'$ so as not to have too light a $W'$.  In order to allow for larger gauge couplings, we need to shrink the size of the conformal window.}
\label{table:model2}
}

\TABLE[t]{
\begin{tabular}{c|c|c|c}
& Model 3a & Model 3b & Model 3c\\
\hline
$L_1$ & $(4.2\TeV)^{-1}$ & $(4.0\TeV)^{-1}$ & $(3.7\TeV)^{-1}$\\
$\epsilon$ & $10^{-8}$ &$10^{-10}$ &$10^{-12}$\\
$g_{EW}$& 0.63 & 0.63  & 0.63\\
$v_{EW}$ & $246 \GeV$ & $246 \GeV$  & $246 \GeV$ \\
\hline
$g'$& $4 g_{EW}$ & $ 3.5 g_{EW}$ & $ 3.0 g_{EW}$ \\
$g_\rho$ &10.4 & 10.1 & 9.2\\
$f_{\pi}$ &$800 \GeV$ & $800 \GeV$ & $800 \GeV$\\
$m_\rho$&  $9.8 \TeV$ & $9.5 \TeV$  & $8.7 \TeV$\\
\hline
$\lambda_{\rm top}$ & 1.0 & 1.0 & 1.0\\
$\nu $&  $+0.49$ & $+0.48$ & $+0.48$\\
$\nu^c$ &  $-0.49$ & $-0.48$ & $-0.48$\\
$\lambda$ &  0.08 & 0.08 & 0.10\\
\hline
$m_{W'}$& $1.0 \TeV$ & $850 \GeV$ & $730 \GeV$\\
$m_{t'}$  & $1.6 \TeV$ &$1.5 \TeV$ &$1.5 \TeV$\\
$m_{q'}$  & $1.6 \TeV$ &$1.5 \TeV$ &$1.5 \TeV$\\
\hline
$m_{h_0}$  &  $210 \GeV$ &  $210 \GeV$ &  $190 \GeV$\\
$\delta_{\rm gauge} m^2_h$  & $+(110 \GeV)^2$ & $+(130 \GeV)^2$  & $+(100 \GeV)^2$\\
$\delta_{\rm fermion} m^2_h$&  $-(290\GeV)^2$ &  $-(300\GeV)^2$ & $-(270 \GeV)^2$\\
$\delta_{\rm other} m^2_h$ &  $+(220 \GeV)^2$ &  $+(220 \GeV)^2$ & $+(200 \GeV)^2$\\
$\delta_{\rm gauge}\lambda_h$  & 0.15  & 0.14 & 0.12 \\
$\delta_{\rm fermion}\lambda_h$ & 0.22  & 0.22 & 0.19\\
\hline
$m_{\phi}$  & $1.3 \TeV$ & $1.1 \TeV$ & $900 \GeV$ \\
$\lambda_{\phi hh}$ & $930 \GeV$& $630 \GeV$ & $360 \GeV$\\
\end{tabular}
\caption{Parameters and results for Model 3.  In this model, we try to equalize the contributions to the Higgs potential from $\delta_{\rm gauge} m^2_h$, $\delta_{\rm fermion} m^2_h$, and $\delta_{\rm other} m^2_h$.  The pion decay constant is fixed at $f_\pi = 800 \GeV$.  The conformal window is smaller than Model 1 but larger than Model 2, and there is more freedom in the choice of $g'$. In order for the fermionic contribution to the Higgs potential to be small, we adjusted $\nu$ and $\nu^c$ such that $t'$ and $q'$ were as light as possible while still allowing $\lambda_{\rm top} = 1$. Note that as we decrease $\epsilon$, we are forced to decrease $g'$, so the $W'$ mass decreases.}
\label{table:model3}
}

The goal of Model 2 is to see if it is possible to reduce the amount of fine-tuning and balance the gauge and fermion contributions.  Once the brane separation $\epsilon$ is chosen, we have very little freedom in the gauge sector if we want $g_\rho$ to be perturbative.   However, simply reducing the scale of the UV brane will not help very much. The real problem is the numerically large difference between the gauge and fermion contributions mentioned above, and short of raising $g_{EW}$ or lowering $\lambda_{\rm top}$, our only freedom is to raise the mass of the $W'$, but this can only get us so far.  In order to dispense with the $\delta_{\rm other} m_h^2$ piece altogether, we have to bring the overall contributions down, and this can be accomplished by lowering the pion decay constant $f_\pi$.  Of course, doing so will cause the $W'$ to be unacceptably light unless we also raise the value of $g'$.  By decreasing the brane separation to make $\epsilon$ larger than $10^{-15}$, we can crank the gauge coupling to $g' = (\mbox{a few})g_{EW}$ while still keeping a perturbative $g_\rho$.  Table \ref{table:model2} presents the parameters of Model 2 where no $\delta_{\rm other} m_h^2$ piece is included.   In Model 2a, we see that simply increasing $\epsilon$ yields a very light $W'$.  Model 2b is much safer, though $g_\rho$ is approaching the edge of the perturbative regime.  In Model 2c, the conformal window is very small, and we see that there is no problem having reasonably heavy partners without including a $\delta_{\rm other} m_h^2$ contribution to artificially raise $f_\pi$ as long as the conformal window is small enough.

Finally, Model 3 represents a compromise between the competing tensions of the theory.  We allow a $\delta_{\rm other} m_h^2$ piece to enforce $f_\pi = 800 \GeV$, but we shrink the conformal window to allow $g' \sim 4 g_{EW}$, effectively putting the UV brane at the intermediate scale.  To have the lightest $q'$ and $t'$ possible, we push $\nu$ and $-\nu^c$ to the edge of the region where we can satisfy the condition $\lambda_{\rm top} = 1$.  This model comes the closest to realizing the original vision of the littlest Higgs, in that the value of $f_\pi$ does not involve an unacceptable level of fine-tuning, the $q'$, $t'$ and $W'$ partners are around $1 \TeV$ in mass, and the scale $m_\rho$ where we see resonances of the strong dynamics is near $10 \TeV$.  Note that $g_\rho \sim 10$ in this model, so we are imagining a very small $N$ CFT.  We see that the mass of the Higgs is around $200 \GeV$, though this value increases substantially if we reduce $g'$ or $\nu$.

In summary, the littlest Higgs in AdS space has a healthy phenomenology if we allow ourselves maximal freedom to raise the mass of the $W'$, lower the mass of the $q'$ and $t'$, and include some (unknown) sources of additional $SU(5)$ violation to get a less negative doublet mass value.  In CFT language, we see that the favored regions of parameter space are small $N$ theories with small conformal windows, though with some amount of fune-tuning we can still have small $g_\rho$ and $\epsilon$.  Note that these tensions are numerical tensions and not parametric tensions.  If we could lower $\lambda_{\rm top}$ and $g_{EW}$, then we could have an exceptionally light Higgs boson with no fine-tuning.  The extent to which these dimensionless parameters are large are the degree to which we have to work to enforce a large separation between the pion decay constant and the electroweak scale.

\section{Toward a Realistic Model with $T$-Parity}
\label{sec:tparity}

To construct a realistic little Higgs model, we need to include the entire fermion sector of the standard model.  One construction is to simply introduce the remaining fermions on the IR brane and include explicit $SO(5)$ violating couplings between these fermions and the Higgs doublet.  This would give rise to a divergent Higgs mass sensitive to the local Planck scale (\emph{i.e.}\ $1/L_1$), but because the non-top Yukawa couplings are sufficiently small, these quadratic divergences would not spoil the successes of the littlest Higgs.  While this construction is consistent, it is not particularly elegant.  Not only does it not explain the hierarchy in the Yukawa couplings, but it does not yield a manifestly finite Coleman-Weinberg potential.

In this section, we sketch one possible AdS implementation of the littlest Higgs where all the fermions live in the bulk of AdS, and the interactions on the IR brane preserve the $SO(5)$ symmetry protecting the Higgs mass.  Just like in Section \ref{section:collective}, the Higgs potential is generated only through loops that stretch from the IR brane to the UV brane.  We would also like to take advantage of some of the previous successes of AdS model building; for example, if the bulk fermions have different bulk masses, we can naturally generate a hierarchy in the Yukawa couplings while simultaneously suppressing flavor changing neutral currents  \cite{Grossman:1999ra,Gherghetta:2000qt,Huber:2000ie,Huber:2003tu}.  As we will see, however, there is a tension between Yukawa coupling hierarchies and $T$-parity.

Given this tension, one may wonder why we want to implement $T$-parity in the first place.  A main constraint on little Higgs theories has been corrections to precision electroweak observables coming from couplings between standard model fields and the new massive gauge fields and scalars in the little Higgs \cite{Hewett:2002px,Csaki:2002qg,Csaki:2003si,Gregoire:2003kr,Kilic:2003mq,Chen:2003fm,Yue:2004xt}.   If these new bosons have mass around $1 \TeV$, then they generate dimension six standard model operators through tree-level exchange suppressed by the scale $1 \TeV$.  However, precision electroweak data suggests that the natural suppression scale should be closer to $5 - 10 \TeV$ \cite{Barbieri:1999tm}.    The difference between the mass scale necessary to stabilize the electroweak scale ($1 \TeV$) and the mass scale suggested by precision electroweak data ($10 \TeV$) is known as the ``little hierarchy'' problem.  In AdS$_5$ language, there is no symmetry that forbids the IR brane operator
\be
\sqrt{-g_\mathrm{ind}} \delta(z-L_1)  \tr \left( F_{\mu \nu} \Sigma F^{*\mu\nu} \Sigma^* \right),
\ee
and when electroweak symmetry is broken, this operator will yield a contribution to the $S$ parameter of order $(v_{EW}/f_\pi)^2$.   As we have seen, while there is a parametric separation between $v_{EW}$ and $f_\pi$, numerically the scales can be close and $S$ may be dangerously large.

In models with $T$-parity \cite{Cheng:2003ju,Cheng:2004yc,Low:2004xc}, there is $\mathbf{Z}_2$ symmetry under which the standard model fields (including the Higgs) are even but the new massive gauge fields and scalars are odd.  Therefore, tree-level exchange of non-SM bosons is forbidden and the dimension six operators generated by integrating out non-SM bosons are suppressed by a loop factor.  $T$-parity is a simple solution to the little hierarchy problem, and generically little Higgs theories with $T$-parity are safe from excessive precision electroweak corrections.  Of course, another way to avoid precision electroweak constraints is to consider a little Higgs mechanism with a custodial $SU(2)$ symmetry \cite{Chang:2003un,Chang:2003zn}, and it would be interesting to try to implement such little Higgs theories in AdS space.  

%In the original $T$-parity formulation of the $SU(5)/SO(5)$ littlest Higgs \cite{Cheng:2004yc}, fermions transformed non-linearly under the $SU(5)$ global symmetry, making it difficult to imagine the bulk fermion content of an AdS extension.  In particular, in a na\"{i}ve implementation of $T$-parity, one would expect to always generate an even number of electroweak doublets.  This is because any bulk fermion charged under $SU(5)$ contains an even number of electroweak doublets, and $T$-parity imposes a $Z_2$ symmetry between $SU(2)_1$ and $SU(2)_2$ such that electoweak doublets must be introduced in pairs on UV brane.  While it is in principle possible to introduce a single electroweak doublet on the IR brane, this doublet would transform non-linearly under $SU(5)$, requiring some CCWZ gymnastics to couple the IR brane doublet to bulk $SU(5)$ fermions. --- As we will see, in this model, it is possible to have an odd number of $SU(2)$ doublets without having to consider non-linear representations of $SU(5)$.

In the original $T$-parity formulation of the $SU(5)/SO(5)$ littlest Higgs \cite{Cheng:2004yc}, fermions transformed non-linearly under the $SU(5)$ global symmetry, making it difficult to imagine the bulk fermion content of an AdS extension.   In  \cite{Low:2004xc}, the $SU(5)/SO(5)$ littlest Higgs was extended to an $SU(5)^2/SO(5)$ model in which all fermions transform linearly under $SU(5)^2$ symmetry, making it much easier to imagine a UV completion of the low energy effective theory.   One difficulty of implementing a realistic $SU(5)^2/SO(5)$ model in AdS space (or even a $SU(5)/SO(5)$ without $T$-parity), is the necessity gauging $U(1)_B$ in the bulk.  Because of this $U(1)_B$, we are forced to introduce a top-type and a bottom-type quark doublet in the bulk, and then use boundary conditions on the UV brane to identify a linear combination of the two doublets as the standard model quark doublet.  Therefore, this model has a large number of $SU(5)$ muliplets to allow for $T$-parity and the $U(1)_B$ nuisance, and it may be interesting to see the effect of such a large number of bulk fermions on the $SU(3)_C$ and $SU(2)_{EW}$ beta functions to see whether we maintain perturbative gauge couplings.

The bulk symmetry in this model is $SU(5)_L \times SU(5)_R$ which is broken to a diagonal $SO(5)_V$ on the IR brane.  We can characterize this breaking pattern by the vacuum expectation values of three scalars.  Imagine three $5\times5$ matrix fields $\Phi_L$, $\Phi_R$, and $\Phi_T$ on the IR brane that transform under $SU(5)_L \times SU(5)_R$ as
\be
\label{phitransform}
\Phi_L \rightarrow L \Phi_L L^T, \qquad \Phi_R \rightarrow R \Phi_R R^T, \qquad \Phi_T \rightarrow L \Phi_T R^\dagger,
\ee
where $L \in SU(5)_L$ and $R \in SU(5)_R$. These fields take vacuum expectation values
\be
\langle \Phi_L \rangle = \Sigma_0, \qquad \langle \Phi_R \rangle = \Sigma_0, \qquad \langle \Phi_T \rangle = \openone.
\ee
We see readily that the vevs of $\Phi_i, i= L, R$ break $SU(5)_i$ to $SO(5)_i$, and the vev of $\Phi_T$ selects the diagonal subgroup of $SO(5)_L \times SO(5)_R$.

Let $T_i^a$ be the generators of $SO(5)_i$ and $X_i^a$ be the generators of $SU(5)_i/SO(5)_i$.  The Goldstone matrices of $SU(5)^2/SO(5)$ can be parametrized as
\be
\Pi_L = \pi^a_L X^a_L,\qquad \Pi_R = \pi^a_R X^a_R,\qquad \Pi_T = \pi^a_T (T^a_L - T^a_R).
\ee
Performing the broken symmetry on the vacuum, the CCWZ prescription \cite{Coleman:1969sm,Callan:1969sn} tells us that
\be
\Sigma_L = e^{2i\Pi_L/f_5} \Sigma_0,\qquad \Sigma_R = e^{2i\Pi_R/f_5} \Sigma_0, \qquad \Sigma_T = e^{i\Pi_L/f_5} e^{2i \Pi_T/f_5} e^{-i \Pi_R/f_5},
\ee
transform like their counterparts in equation (\ref{phitransform}).  For simplicity, we have given each $\Sigma$ field the same 5D decay constant $f_5$, which as before will be taken to infinity in any calculation.

In addition to introducing an $SU(3)_C \times U(1)_B$ bulk gauge symmetry, we also gauge a $SU(2)_{L1} \times SU(2)_{L2} \times SU(2)_R \times U(1)_Y$ subgroup of $SU(5)^2$, \emph{i.e.}\ we give this subgroup Neumann boundary conditions on the UV brane.\footnote{In order to give $U(1)_Y$ a different gauge coupling from $SU(2)_{Li}$ we also introduce boundary gauge kinetic terms.}  The subgroup is imbedded as
\be
Q_{L1}^a =\left(\begin{array}{cc}\sigma^{a}/2 & \quad \\ \quad&\quad \end{array}\right)_L, \qquad Q_{L2}^a = \left(\begin{array}{cc} \quad&\quad  \\ \quad &-\sigma^{*a}/2\end{array}\right)_L, \qquad Q_{R}^a = \left(\begin{array}{cc}\sigma^{a}/2 \quad&\quad  \\ \quad &-\sigma^{*a}/2\end{array}\right)_R,
\ee
\be
A = \frac{1}{2}\diag(1,1,0,-1,-1)_L,
\ee
where subscripts $L$ and $R$ on the matrices indicate whether the gauge generator belongs to $SU(5)_L$ or $SU(5)_R$.  The hypercharge generator is $Y = A + B$, and when $SU(5)^2$ breaks to $SO(5)$, the electroweak $SU(2)$ is generated by
\be
Q^a_{EW} = Q_{L1}^a+Q_{L2}^a+Q_{R}^a.
\ee
Note that the same subgroup of $SU(5)_L$ is gauged as in Section \ref{section:model}.

We identify the Goldstone matrix $\Pi_L$ with the Goldstone matrix in the original $SU(5)/SO(5)$ little Higgs.  As shown in \cite{Low:2004xc} the Goldstones in $\Pi_R$ and $\Pi_T$ are either eaten by the broken $SU(2)$ gauge  bosons or can be given large masses through radiatively generated gauge interactions or through plaquette operators.  These plaquette operators live on the IR brane and explicitly break $SU(5)^2$ even before the $\Phi$ fields take their vaccum expectation value.  However, they do so in a way that maintains the $SU(2)^3$ subgroup of $SU(5)^2$, so the low energy gauge structure is not modified.

In the fermion sector,  the action of $T$-parity effectively maps a $\mathbf{\bar{5}}$ of $SU(5)_L$ onto a $\mathbf{5}$ of $SU(5)_L$ and leaves representations of $SU(5)_R$ unchanged.  One linear combination of the $\mathbf{\bar{5}}_L$ and $\mathbf{5}_L$ becomes a standard model fermion, and a $\mathbf{5}_R$ marries the other combination to become the heavy $T$-odd partner to the standard model fermion.  Therefore, for each $SU(5)$ multiplet $\Psi$ corresponding to a standard model fermion, we introduce three bulk Dirac fermions $\mathbf{\Psi}_{L1}$, $\mathbf{\Psi}_{L2}$, and $\mathbf{\Psi}_{R}$.  We choose these fields to have non-vanishing upper (left-handed) components on the IR brane. These upper component Weyl fields $\Psi_{L1}$, $\Psi_{L2}$, and $\Psi_{R}$ transform under $SU(5)_L \times SU(5)_R$ as:
\be
\label{eqn:psiconventions}
\begin{array}{l|ccc}
& SU(5)_L & SU(5)_R & \mbox{Other Quantum Numbers}\\
\hline
\Psi_{L1} & \mathbf{\bar{5}}  &  - & \mbox{Defined by }\Psi \\
\Psi_{L2} & \mathbf{5}  &  - & \mbox{Defined by }\Psi \\
\Psi_{R}  & - &  \mathbf{5} & \mbox{Conjugate to }\Psi\\
\end{array}
\ee
In the last column we mean that $\Psi_{L1}$ and $\Psi_{L2}$ have identical $SU(3)_C \times U(1)_B$ quantum numbers and $\Psi_R$ has the conjugate quantum numbers.  For convenience, we also introduce the notation for bulk Dirac fermions $\mathbf{\Psi}^c_{L1}$, $\mathbf{\Psi}^c_{L2}$, and $\mathbf{\Psi}^c_{R}$ whose lower (right-handed) components are non-vanishing on the IR brane.    These lower component Weyl fields $\Psi^c_{L1}$, $\Psi^c_{L2}$, and $\Psi^c_{R}$ transform under $SU(5)_L \times SU(5)_R$ the same way as $\Psi_{L1}$, $\Psi_{L2}$, and $\Psi_{R}$ but have opposite other quantum numbers to their counterparts.
\be
\label{eqn:psicconventions}
\begin{array}{l|ccc}
& SU(5)_L & SU(5)_R & \mbox{Other Quantum Numbers}\\
\hline
\Psi^c_{L1} & \mathbf{\bar{5}}  &  - & \mbox{Conjugate to }\Psi \\
\Psi^c_{L2} & \mathbf{5}  &  - & \mbox{Conjugate to }\Psi \\
\Psi^c_{R}  & - &  \mathbf{5} & \mbox{Defined by }\Psi\\
\end{array}
\ee

We now define the action of $T$-parity.  Start with the $5\times5$ matrices
\be
\Omega = \left(\begin{array}{ccc}-\openone &  &  \\ & 1 &  \\ &  & -\openone\end{array}\right), \qquad Z = \Sigma_0 \Omega, \qquad Z^2 = \Omega^2 = \openone.
\ee
Let $A_i$ be the $SU(5)_i$ gauge fields for $i=L,R$.   The action of $T$-parity looks like charge conjugation on $SU(5)_L$.
\be
\label{eqn:Tgauge}
A_L \rightarrow -Z A_L^T Z,\qquad A_R \rightarrow \Omega A_R \Omega.
\ee
In terms of the gauge fields on the UV brane, $T$-parity maps $SU(2)_{L1}$ to $SU(2)_{L2}$ and leaves $SU(2)_R$ and $U(1)_Y$ invariant.  Note that $T$-parity forces the gauge couplings $g_{L1}$ and $g_{L2}$ to be equal.  Before $SU(2)^3$ is broken to the electroweak $SU(2)$, there are two $T$-even gauge bosons and one $T$-odd gauge boson.
\be
Q^a_{EW} = Q_{L1}^a+Q_{L2}^a+Q_{R}^a, \qquad Q^a_+ = Q_{L1}^a+Q_{L2}^a - 2 Q_{R}^a, \qquad Q^a_- = Q_{L1}^a - Q_{L2}^a.
\ee
After $SU(2)^3$ breaks to $SU(2)$, $Q^a_\pm$ get masses.  $Q^a_-$ has a mass $g_L f_4 \sim 1 \TeV$, and by $T$-parity it cannot contribute to tree-level dimension six operators.  However, $Q^a_+$ can contribute to tree-level dimension six operators, so we must choose $SU(2)_R$ to have a large $g_R \sim 4\pi$ gauge coupling in order for $Q^a_+$ to get a mass of order $g_R f_4 \sim 10 \TeV$.  Also, this forces $Q^a_+$ to be mostly $Q^a_R$, and if standard model fields are uncharged under $SU(2)_R$, then precision electroweak corrections from tree-level $Q^a_+$ exchange is suppressed.  Note that in order to really have a large $g_R$, we would have to have a very small conformal window.

On the Goldstone matrices, $T$-parity acts as
\be
\Sigma_L \rightarrow Z \Sigma_L^\dagger Z,\qquad \Sigma_R \rightarrow \Omega \Sigma_R \Omega, \qquad \Sigma_T \rightarrow Z \Sigma_L  \Sigma_T \Omega.
\ee
The action of $T$-parity on $\Sigma_T$ is indeed a $\mathbf{Z}_2$ symmetry.  On the fields in $\Pi_L$ as defined in (\ref{eqn:pifield}), $h$ is $T$-even, and $\eta$ and $\phi$ are $T$-odd, therefore $\phi$ and $\eta$ can safely have $1 \TeV$ masses without affecting precision electroweak data.  Also, it is a quick check that the Goldstone kinetic terms are invariant under $T$-parity.

Finally, the action of $T$-parity on the fermions must be consistent with the gauge sector in equation (\ref{eqn:Tgauge}).  Again, $T$-parity looks like charge conjugations between a  $\mathbf{\bar{5}}_L$ ($\mathbf{\Psi}_{L1}$) and a $\mathbf{5}_L$ ($\mathbf{\Psi}_{L2}$):
\be
\mathbf{\Psi}_{L1} \rightarrow Z  \mathbf{\Psi}_{L2},\qquad \mathbf{\Psi}_{L2} \rightarrow Z  \mathbf{\Psi}_{L1}, \qquad\mathbf{\Psi}_{R} \rightarrow \Omega  \mathbf{\Psi}_{R},
\ee
with similar formulas for the $\Psi^c$ fermions.  Before introducing the specific quantum numbers of the standard model fields, note that we can give mass to the $T$-odd combination of $\Psi_{L1}$ and $\Psi_{L2}$ via the interaction on the TeV brane:
\be
\label{mirrormass}
\mathcal{L}_{T\mathrm{-odd~mass}}=\sqrt{-g_\mathrm{ind}}\delta(z-L_1)  \kappa \left(\Psi_{L1} \Sigma_T \Psi_R + \Psi_{L2} \Sigma_L \Sigma_T \Psi_R + h.c. \right).
\ee
We can write down a similar interaction for $\Psi^c$.   Because $\Sigma_T$ depends on $\Pi_L$, one might worry that this interaction could give rise to a radiative correction the the Higgs mass.  In the case that $\Psi$ is supposed to describe an electroweak doublet,  $\Psi_{L1}$, $\Psi_{L2}$, and $\Psi_{R}$ have boundary conditions that happen to preserve an $SU(3)^2$ symmetry that is enough to protect the Higgs mass.  If $\Psi$ describes an electroweak singlet however, there is no symmetry protecting the Higgs mass and there will be a radiatively generated potential. Just as for the top sector from Section \ref{section:collective}, though, the interaction in equation (\ref{mirrormass}) exhibits collective breaking to the extent that a potential is only generated if both $\Psi_{Li}$ and $\Psi_R$ have $SU(5)$ violating boundary conditions.

The $T$-invariant Yukawa interactions between $\Psi$, $\Psi^c$, and the Higgs take the form
\be
\label{eqn:tyukawa}
\mathcal{L}_\mathrm{Higgs} =\sqrt{-g_\mathrm{ind}} \delta(z-L_1) \lambda \left(\Psi_{L1} \Sigma_L \Psi^c_{L1}+\Psi_{L2} \Sigma_L^\dagger \Psi^c_{L2} + h.c. \right).
\ee
This is simply the $T$-invariant generalization of equation (\ref{eqn:yukawainteraction}).  Already, we can see the tension between Yukawa coupling hierarchies and the masses of the $T$-odd fermion partners.  If $\lambda$ is $\mathcal{O}(1)$, we know we can generate the Yukawa coupling hierarchy through the different overlaps of  $\Psi_L$ and $\Psi^c_L$ with the IR brane for each generation.  However, the same overlap functions would enter into equation (\ref{mirrormass}).  We could of course make  $\kappa$ large, insist that $\Psi_R$ and $\Psi^c_R$ have maximum overlap with the IR brane, and also split the Yukawa hierarchy equally between the $\Psi_L$ and $\Psi^c_L$ overlaps.  But given that the electron mass is a factor of $10^6$ lighter than the top quark, we would find that the $T$-odd partner to the electron is $\sqrt{10^6} = 10^3$ lighter than the $T$-odd partner of the top quark!   We will discuss this issue further at the end of this section.

We now give the quantum numbers of a standard model generation.  As mentioned already, there are slight complications coming from the fact that we need to gauge $U(1)_B$.  This forces us to introduce a top-type doublet and a bottom-type doublet.  Ignoring a possible right-handed neutrino, we have the following $SU(5)_L \times SU(5)_R$ matter content, using the notation of equations (\ref{eqn:psiconventions}) and (\ref{eqn:psicconventions}):
\be
\begin{array}{l|cc}
&SU(3)_C & U(1)_B\\
\hline
U& \mathbf{3} & +2/3\\
U^c& \mathbf{\bar{3}} & -2/3\\
\hline
D& \mathbf{3} & -1/3\\
D^c& \mathbf{\bar{3}} & +1/3\\
\hline
L& - & - 1\\
L^c & - & +1\\
\end{array}
\ee
Looking at the top sector in detail, we imbed the top-type doublet and top singlet fields as
\be
\label{eqn:uinfo}
U_{L1} = \left(\begin{array}{c} q_1^U \\ - \\ - \end{array}\right), \qquad U_{L2} = \left(\begin{array}{c} - \\ - \\ q_2^U \end{array}\right), \qquad U_{R} = \left(\begin{array}{c} q_M^U \\ - \\ - \end{array}\right),
\ee
\be
\label{eqn:ucinfo}
U^c_{L1} = \left(\begin{array}{c} - \\ u_1^c \\ - \end{array}\right), \qquad U^c_{L2} = \left(\begin{array}{c} - \\ u_2^c \\ - \end{array}\right), \qquad U^c_{R} = \left(\begin{array}{c} - \\ u_M^c \\ - \end{array}\right),
\ee
where dashes indicate fields that have Dirichlet boundary conditions on the UV brane and therefore do not have zero modes.  As in Section \ref{section:Low Energy Matching}, the physical top doublet and singlet will be mixtures of these zero modes and components from the KK modes.  After $SU(2)^3$ breaks to $SU(2)_{EW}$, the $q^U$ and $u^c$ fields have the right standard model quantum numbers.  At low energies, the interaction in equation (\ref{mirrormass}) generates the mass terms
\be
\mathcal{L}_\mathrm{mass} \sim  m^U_q \; q_M^U(q_1^U + q_2^U) +  m_u \; u_M(u_1 + u_2),
\ee
so at low energies, the fields
\be
q^U_\mathrm{even} = \frac{1}{\sqrt{2}}(q_1^U - q_2^U), \qquad u^c_\mathrm{even} = \frac{1}{\sqrt{2}}(u^c_1 - u_2^c),
\ee
are massless, $T$-even fermions, and the orthogonal $T$-odd combinations get masses $m^U_q$ and $m_u$.  At energies below the $T$-odd masses, the interaction in equation (\ref{eqn:tyukawa}) generates the Yukawa coupling
\be
\mathcal{L}_\mathrm{Yukawa} \sim \lambda^U \, q^U_\mathrm{even} h u^c_\mathrm{even}.
\ee
As in Section \ref{section:model}, the physical Yukawa coupling involves wavefunction renormalization from mixing with heavy KK modes.

By this prescription, we will have a massless $q^U_\mathrm{even}$ and a massless $q^D_\mathrm{even}$ both with the quantum numbers of a quark doublet.   We want to identify one combination of $q_\mathrm{even}^U$ and $q_\mathrm{even}^D$ to be the standard model quark doublet and the other combination to be heavy.  Looking at the $D$ field:
\be
D_{L1} = \left(\begin{array}{c} - \\ - \\ q_2^D \end{array}\right), \qquad D_{L2} = \left(\begin{array}{c} q_1^D \\ - \\ - \end{array}\right), \qquad D_{R} = \left(\begin{array}{c} - \\ - \\ q_M^D \end{array}\right).
\ee
(The subscripts $1$, $2$ on $q_1^D$ and $q_2^D$ indicate whether the field transforms under $SU(2)_{L1}$ or $SU(2)_{L2}$, and \emph{not} whether they came from the field $D_{L1}$ or $D_{L2}$.)  Consistent with $SU(2)_i$, we can set boundary conditions on the UV brane for $(q^U_1 - q^D_1)$, $(q^U_2 - q^D_2)$, and $(q^U_M - q^D_M)$ to vanish. The $(q^U_M - q^D_M)$ boundary condition is necessary because without it, there would be an additional massless doublet.  At the end of the day, the combination
\be
q_\mathrm{even} = \frac{1}{\sqrt{2}}(q^U_\mathrm{even} + q^D_\mathrm{even})
\ee
is the massless standard model quark doublet and it couples in the appropriate way to $u^c$ and $d^c$.  Again, the physical standard model fermions will be mixtures of these $T$-even zero modes with components of $T$-even KK states.  In the lepton sector, we imbed the lepton doublet and singlet analogously to equations (\ref{eqn:uinfo}) and (\ref{eqn:uinfo}), but if we do not have a right-handed neutrino, we do not need to have a separate electron-type and neutrino-type lepton doublet.

While it might be interesting to calculate the Higgs radiative potential in this $T$-parity extension of the AdS$_5$ littlest Higgs, there is a basic model building question that needs to be addressed before such a calculation would become meaningful.  As we saw in Section \ref{section:collective}, the Higgs potential depends on the bulk propagators for the fermions, and these in turn depend heavily on the choice of fermion bulk masses.  If we simply want to reproduce the Higgs Yukawa matrix and ignore flavor-violating four-fermion operators, then we can simply choose some reasonable fermion bulk masses that give $\mathcal{O}(1)$ overlap with the IR brane and then dial the Yukawa matrix by hand on IR brane to get the desired low energy Yukawa structure.  In this case, the mass of \emph{all} the $T$-odd fermions can be pushed up to around $10 \TeV$ with a reasonable value of $\kappa$ in equation (\ref{mirrormass}).  (See \cite{Low:2004xc} for a reason why the $T$-odd fermions most likely need to be closer to $1 \TeV$.)

If, however, we want to naturally explain the smallness of four-fermion operators that contribute to FCNCs, then we want the different fermion generations to be localized in different parts of the bulk.  This means that there will be a hierarchy in the overlap of the different generations with the IR brane, and while this feature is desirable from the point of view of trying to understand the Yukawa hierarchy, it is disastrous in light of  equation (\ref{mirrormass}), where the overlap functions also determine the masses of the $T$-odd fermions.  We expect this to be a general issue with trying to build AdS models of the little Higgs with $T$-parity.  The problem is not with $T$-parity itself; indeed, for a single generation where flavor is not an issue we can easily incorporate $T$-parity in AdS space.  The problem is that $T$-parity wants to be flavor blind (\emph{i.e.}\ all of the $T$-odd fermions need to have roughly the same mass), whereas our implementation of $T$-parity makes explicit reference to flavor because the $T$-odd mass terms inherit flavor specific overlap functions.  We may have some freedom to generate the Yukawa hierarchy through a combination of overlap functions and IR brane matrix elements, but in the context of AdS model building, there is no explanation for why these two effects would naturally work in the same direction.  There may be other implementations of $T$-parity in AdS space that avoid these issues, but in this model there is an important tension between $T$-parity and flavor.

\section{Future Directions}

We have seen that the collective breaking structure of little Higgs theories can be naturally implemented in AdS space.  The 5D Coleman-Weinberg potential has a particularly illuminating structure, in that corrections to the Higgs potential are proportional to \emph{two} differences of 5D Greens functions.  In this paper, we have worked out the gauge and fermion structure for the $SU(5)/SO(5)$ littlest Higgs, but the construction can be easily extended to other little Higgs theories.  The AdS/CFT correspondence tells us exactly how to implement the gauge and Goldstone sector of any $G / H$ little Higgs theory.  The only possible challenge is trying to figure out the bulk $G$ fermion content, and in general it is desirable to start with a low energy theory where the fermions come in linear representations of $G$.

From the top-down point of view, we have found a UV completion of the littlest Higgs that concisely explains the hierarchy between the Planck scale and electroweak scale.  We start at some high scale with a large $N$ conformal field theory.  The size of AdS space represents the logarithmic running of some operator which eventually breaks conformality in the infrared, and this guarantees a natural hierarchy between $M_\Pl$ and the $10 \TeV$ scale.   Because conformality is broken, the CFT confines, yielding an $SU(5)/SO(5)$s worth of ``pions.''   Collective breaking insures that the pion corresponding to the Higgs boson is light, and this generates a natural hierarchy between $f_\pi \sim 1 \TeV$ and the electroweak scale.

We have seen a number of tensions in this model that one could guess from the low energy theory.  Large $N$ CFTs with a large conformal window lead to very small gauge couplings, and if we want to raise the mass of the $W'$ to phenomenologically healthy values, we need to shrink the size of the conformal window.  If we allow for $T$-parity, then a light $W'$ may be acceptable, but in the context of AdS space, we saw a tension between flavor physics, precision electroweak tests, and little Higgs theories.  If $T$-parity is indeed the reason for the smallness of precision electroweak corrections, then we have to explain why the masses of $T$-odd fermion partners do not exhibit the same flavor hierarchy as the fermions themselves.  It may be interesting to look at AdS models of little Higgs theories with custodial $SU(2)$ symmetry where it appears easier to separate flavor physics from precision electroweak tests.

In AdS space, locality in the warped dimension guaranteed that our Higgs potential was finite, but one might wonder how important the extra dimension actually was for our construction to succeed.   Even if we took $\epsilon = L_0 / L_1$ to be $1/2$, the Higgs potential would still be finite, suggesting that the physics relevant for pseudo-Goldstone phenomenology is heavily localized near the IR brane.  In fact, as shown in \cite{Piai:2004yb}, the Higgs potential can be made completely free of 1-loop quadratic divergences if one simply postulates the existence of $\rho$-like mesons.  Of course, the properties of the $\rho$ are inherited from the strong dynamics, and AdS/CFT is a simple way to understand $\rho$-like states.   But given the difficulty of having heavy $W'$ states in large $N$ CFTs, it would be nice if we had a framework other than AdS/CFT to understand  the radiative potentials for pseudo-Goldstone bosons.

In \cite{Thaler:2005kr}, we will show that in the right context, pseudo-Goldstone phenomenology can be made largely independent of strong dynamics.  In essence, a low energy observer cannot tell whether a gauge quadratic divergence is cut off by a $\rho$-like or a $W'$-like state, so if our only problem with a large conformal window is that $W'$ states are generically much lighter than $\rho$ states, then we should work in a framework where we can decouple the $\rho$-like states.  In particular, we will show that the $SU(5)/SO(5)$ littlest Higgs can actually arise in ordinary QCD with five flavors, and the gauge contribution to the Higgs potential is completely finite at one-loop.  This UV complete theory will have light $W'$ and $W''$ states responsible for cutting off divergences in the Higgs potential, but the QCD $\rho$ mesons will play a much smaller role in the low energy phenomenology.   This will open a new avenue to construct simple UV completions of little Higgs models.

\acknowledgments{We thank Nima Arkani-Hamed for inspiring this project and providing helpful guidance throughout this work.  We gratefully acknowledge Spencer Chang,  Hsin-Chia Cheng, Can Kilic, Ian Low, Rakhi Mahbubani, Jay Wacker, and Devin Walker for stimulating conversations on little Higgs model building.}

\appendix

\section{Summary of Bulk Fields in AdS Space}
\label{conventions}

In this appendix, we briefly summarize the main results and conventions for bulk fields in AdS space. All the results below can be found in the literature (see for example \cite{Contino:2003ve, Agashe:2004rs, Grossman:1999ra, Gherghetta:2000qt}), though there are different conventions for the naming of left- and right-handed fermions, and we try to stick to the conventions of \cite{Grossman:1999ra}. Throughout, we use the choice of metric given by equation (\ref{eqn:metric}), which is most convenient for our purposes. 

Let us begin with scalars in the bulk. The lagrangian for a complex scalar field in 5D with arbitrary boundary mass terms and a bulk source $J$ is,
\be
\mathcal{L}_\phi = \sqrt{g}\left( g^{M N} \partial_M \phi^{*} \partial_N \phi - M^2 |\phi|^2 - J \phi\right) + \sqrt{-g_{\mathrm{ind}}} m_0^2 |\phi|^2 \delta (z-L_0)  + \sqrt{-g_{\mathrm{ind}}} m_1^2 |\phi|^2 \delta (z-L_1).
\ee
The induced metric on the branes is $g_{\mathrm{ind}}^{\mu\nu}(z) = \eta^{\mu\nu}/(k z)^2$. The boundary terms are merely a handy way of imposing boundary conditions on the fields in the lagrangian formulation. Rescaling the fields in the lagrangian simplifies things somewhat and allow for a unified treatment of fields with different spins, so let $\phi \rightarrow \hat{\phi} = (kz)^{-2} \phi$. The lagrangian is then (ignoring the boundary terms since they will only change the boundary conditions of the Greens function),
\begin{equation}
\mathcal{L}_\phi = (k z)\left( \left(-\hat{\phi}^*\partial^{\mu} \partial_{\mu}\hat{\phi} + \hat{\phi}^* \left(\partial^2_z + \frac{1}{z}\partial_z - \frac{4}{z^2} \right) \hat{\phi} \right)- \frac{M^2}{(kz)^2}|\hat{\phi}|^2 - \frac{\sqrt{g}}{(kz)}\hat{J} \hat{\phi}\right).
\end{equation}
The rescaled Greens function $\hat{G}(x,x',z,z') = (kz)^{-2} (kz')^{-2} G(x,x',z,z')$ satisfies the following differential equation (moving to momentum space in the transverse direction),
\begin{equation}
\label{eqn:greens}
\left(p^2 + \partial^2_z + \frac{1}{z}\partial_z - \frac{\alpha ^2}{z^2}\right) \hat{G}(p; z,z') = \frac{1}{kz} \delta(z-z'),
\end{equation}
where $\alpha^2 = M^2/k^2 + 4$.

Let us move on to consider gauge bosons in AdS space. Similar to the scalar case, the 5D gauge boson lagrangian is given by
\be
\label{eqn:bulkgauge}
\mathcal{L}_{\rm gauge} = -\frac{1}{4g_5^2} \sqrt{g}g^{AB}g^{MN}F_{AM}F_{BN} - \frac{z_0}{4g_5^2}\sqrt{-g_\mathrm{ind}}g_\mathrm{ind}^{\mu \nu}g_\mathrm{ind}^{\alpha\beta}F_{\mu \alpha}F_{\nu \beta}\delta (z-L_0),
\ee
where we have dropped a possible IR brane kinetic term since it plays no role in our analysis.  Working in $A_5 = 0$ and $\partial_{\mu} A^{\mu} =0$ gauge, the situation is identical to the scalar case, except that the rescaled field is $\hat{A}_{\mu} = (kz)^{-1} A_{\mu}$, and so the rescaled propagator is $\eta_{\mu\nu}\hat{G}(p;z,z') = (kz)^{-1} (kz')^{-1} \eta_{\mu\nu} G(p;z,z')$. The propagator satisfies equation (\ref{eqn:greens}) with $\alpha = 1$.

In curved backgrounds, the spin of a field is defined with respect to the Lorentz group acting on the comoving coordinates (the local tangent space). The vierbein $e_A^a$ is the object which connects the comoving frame ($a$ index) with the spacetime coordinates ($A$ index). The spin-connection $w_{bcA}$ tells us how the comoving reference frame changes as we go in different spacetime directions (hence the two internal indices $a$ and $b$ and one spacetime index $A$). In 5 dimensions, Dirac fermions form an irreducible representation of the Lorentz group. The lagrangian for a Dirac field in the bulk is
\begin{equation}
\label{eqn:bulkfermion}
\mathcal{L}_{\rm fermion} = \sqrt{g} \left( e_a^A\left(\frac{i}{2} \bar{\Psi} \gamma^a (\partial_A - \overleftarrow{\partial}_A)\Psi + \frac{w_{bcA}}{8} \bar{\Psi}\{\gamma^a, \sigma^{bc}\}\Psi \right) - m \bar{\Psi}\Psi - \bar{\Psi}J\right),
\end{equation}
where $e_a^A = \diag(1,1,1,1,1)/kz$ is the inverse vierbein. Since the metric is diagonal, the spin-connection is non-zero only when $b=A$ or $c=A$, giving no contribution to the fermionic action. Again, rescaling $\Psi \rightarrow \hat{\Psi} = (k z)^{-5/2} \Psi$ the lagrangian takes the form,
\begin{equation}
\mathcal{L}_{\rm fermion} = (k z) \left(\hat{\bar{\Psi}} i \slashed{\partial} \hat{\Psi} - \hat{\bar{\Psi}}\left(\frac{m}{kz} + \frac{1}{2z}\gamma_5\right)\hat{\Psi} - \hat{\bar{\Psi}}\gamma_5\partial_z \hat{\Psi} - \frac{\sqrt{g}}{kz} \hat{\bar{\Psi}}J\right).
\end{equation}
Therefore, the rescaled fermion Greens function have to satisfy,
\begin{equation}
\left(\slashed{p} + \frac{\nu}{z} +\gamma_5 \left(\frac{1}{2z} + \partial_z\right)\right) \hat{S}(p;z,z';\nu) = \frac{-1}{kz} \delta(z-z'),
\end{equation}
where we have introduced the ratio $\nu \equiv m/k$.  If we let
\begin{equation}
\hat{S}^{(\pm,\pm)}(p;z,z';\nu) = \left(-\slashed{p} - \gamma_5\left(\partial_z + \frac{1}{2z}\right) + \frac{\nu}{z}\right) (P_L \hat{G}_L^{(\pm,\pm)}+P_R \hat{G}_R^{(\mp,\mp)}),
\end{equation}
we see that $\hat{G}$ has to satisfy equation (\ref{eqn:greens}) again with $\alpha = |1/2 \mp \nu|$ with the minus (plus) sign for the left- (right-) handed propagator.  The $\pm$ signs on the $G$s refer to the choice of boundary conditions for the fields.

So it all comes down to solving equation (\ref{eqn:greens}). The boundary conditions are as follows. We can choose either Neumann (even) or Dirichlet (odd) boundary conditions on the two branes. For the fermions the boundary conditions for the left- and right-handed components have to be opposite from each other since as we will see they are coupled through the equations of motion. The solution to equation (\ref{eqn:greens}) is simple and is given in \cite{Contino:2003ve, Gherghetta:2000kr} (note that we have analytically continued to Euclidean space so that $p^2\rightarrow -p^2$):
\begin{equation}
\label{eqn:generalgreen}
\hat{G}(p;z,z') = \frac{-L_0}{R_1 - R_0} \left(\vphantom{\sum}
I_{\alpha}(|p|z_<) - R_0 K_{\alpha}(|p|z_<)\right)\left(\vphantom{\sum}I_{\alpha}(|p|z_>) - R_1 K_{\alpha}(|p|z_>)\right),
\end{equation}
where $z_>$ ($z_<$) is the greater (lesser) of $(z,z')$, $I_{\alpha}$ and $K_{\alpha}$ are the modified Bessel's functions and the ratios $R_0,R_1$ depend on the choice of boundary conditions on the UV/IR branes. For the gauge boson:
\begin{equation}
\nonumber
R_0^{(+)} = \frac{I_{\alpha-1}(|p|L_0)-z_0|p|I_{\alpha}(|p|L_0)}{-K_{\alpha-1}(|p|L_0)-z_0|p|K_{\alpha}(|p|L_0)},   \qquad R_0^{(-)} = \frac{I_{\alpha}(|p|L_0)}{K_{\alpha}(|p|L_0)},
\end{equation}
\begin{equation}
R_1^{(+)} = \frac{I_{\alpha-1}(|p|L_1)}{-K_{\alpha-1}(|p|L_1)},   \qquad R_1^{(-)} = \frac{I_{\alpha}(|p|L_1)}{K_{\alpha}(|p|L_1)}.
\end{equation}
For the fermions, the ratios are different for the left- and right-handed components of the Dirac field. To simplify the expressions somewhat let  $\alpha = |1/2 \mp \nu|$ and $\beta = 1/2 \mp \nu$:
\begin{equation}
\nonumber
R_0^{(+)} = - \frac{|p|L_0 I_{\alpha-1}(|p|L_0) - \left(\alpha - \beta\right)I_{\alpha}(|p|L_0)}{|p|L_0 K_{\alpha-1}(|p|L_0) + \left(\alpha - \beta\right)K_{\alpha}(|p|L_0)}, \qquad
R_0^{(-)} =\frac{I_{\alpha}(|p|L_0)}{K_{\alpha}(|p|L_0)},
\end{equation}
\begin{equation}
R_1^{(+)} = - \frac{|p|L_1 I_{\alpha-1}(|p|L_1) - \left(\alpha - \beta\right)I_{\alpha}(|p|L_1)}{|p|L_1 K_{\alpha-1}(|p|L_1) + \left(\alpha-\beta\right)K_{\alpha}(|p|L_1)}, \qquad
R_1^{(-)} =\frac{I_{\alpha}(|p|L_1)}{K_{\alpha}(|p|L_1)}.
\end{equation}
Note that when $\alpha = \beta$, this simply reduces to the same ratios as for gauge bosons with no boundary kinetic terms.   We ignore possible boundary kinetic/mass terms for fermions since we do not need them in this paper.

It is important to know the low energy expansion of these propagators with even boundary conditions on the IR bane evaluated at the IR brane, in order to match the UV theory with the IR theory.  For the gauge bosons we have,
\be
\hat{G}^{(+,+)}(p;L_1,L_1) = \frac{\epsilon^2}{L_0 \log(\epsilon^{-1})}\frac{1}{p^2} + \mathcal{O}(p^0), \qquad \hat{G}^{(-,+)}(p;L_1,L_1) = \frac{L_0}{2}(1 + \epsilon^2) + \mathcal{O}(p^2).
\ee
Notice that the expansion of $\hat{G}^{(+,+)}$ is as expected:  it is the gauge boson zero-mode overlap function squared, divided by $p^2$. For the fermions the low energy expansion gives,
\begin{eqnarray}
\hat{G}_L^{(+,+)}(p;L_1,L_1;\nu) &=& \frac{\epsilon}{p^2} \frac{1}{L_1}\left(\frac{1+2\nu}{1-\epsilon^{1+2\nu}} \right)  + \mathcal{O}(p^0)= \frac{\epsilon}{p^2}f_L(\nu)^2,\nonumber\\
\hat{G}_L^{(-,+)}(p;L_1,L_1;\nu) &= &\epsilon L_1 \left(\frac{1-\epsilon^{1-2\nu}}{1-2\nu}\right)  + \mathcal{O}(p^2)= \epsilon \frac{1}{f_R(\nu)^2},\nonumber\\
\hat{G}_R^{(+,+)}(p;L_1,L_1;\nu) &=& \frac{\epsilon}{p^2} \frac{1}{L_1}\left(\frac{1-2\nu}{1-\epsilon^{1-2\nu}} \right)  + \mathcal{O}(p^0)= \frac{\epsilon}{p^2}f_R(\nu)^2,\nonumber\\
\hat{G}_R^{(-,+)}(p;L_1,L_1;\nu) &=& \epsilon L_1 \left(\frac{1-\epsilon^{1+2\nu}}{1+2\nu}\right)  + \mathcal{O}(p^2)= \epsilon \frac{1}{f_L(\nu)^2}.
\end{eqnarray}
Notice that $G_L^{(+,+)}$ ($G_R^{(+,+)}$) is nothing but the left (right) zero-mode overlap function squared divided by $p^2$. It is amusing that $G_L^{(-,+)}$ ($G_R^{(-,+)}$) is simply the inverse of the \emph{right} (\emph{left}) zero-mode overlap function.

The KK spectrum of the theory is found in the poles of the propagator in equation (\ref{eqn:generalgreen}). Since the numerator of the propagator is everywhere analytic, the poles must be at $R_1(p) - R_0(p) =0$. Analytically continuing back to Minkowski space the masses are given by the solutions to
\begin{equation}
 R_1(i m_n) = R_0(i m_n),
\end{equation}
where $R_1$ and $R_0$ for fermions and gauge bosons are given above for the different choices of boundary conditions. For instance, in the case of odd boundary conditions on both branes for a right-handed fermion field we have,
\begin{equation}
\label{eqn: KK mass spectrum}
\frac{J_{\alpha}(m_n L_1)}{Y_{\alpha}(m_n L_1)} = \frac{J_{\alpha}(m_n L_0)}{Y_{\alpha}(m_n L_0)}
\end{equation}
where $\alpha= |1/2+\nu|$. Solving this equation for $m_n$, one obtains the KK spectrum. A similar story holds for gauge boson masses.

Now for the Kaluza-Klein decomposition of the different fields. We will present the details for the fermion KK modes only \cite{Grossman:1999ra}.  Starting from the lagrangian given by equation (\ref{eqn:bulkfermion}) we decompose the 5D Dirac fields into,
\begin{equation}
\Psi_{L,R}(x,z) = \sum_n \psi_n^{L,R}(x,z) (kz)^2 f_n^{L,R}(z).
\end{equation}
To get the usual 4D lagrangian for left- and right-handed fermions we demand that,
\begin{equation}
\label{eqn:fermionnor}
 \int dz (f_n^L)^* f_m^L = \int dz (f_n^R)^* f_m^R = \delta_{m,n},
\end{equation}
\begin{equation}
\label{eqn:fermioneom}
\left( \pm z \partial_z - \nu\right)f_n^{L,R}(z) = - m_n z f_n^{R,L}(z),
\end{equation}
where $m_n$ is the 4-dimensional mass for the different KK modes.
The boundary conditions on $f_n^{L,R}$ are simply $f_n^{L*}(L_0)f_n^{R}(L_0) = f_n^{L*}(L_1)f_n^{R}(L_1) = 0$, which tells us that either the left- or right-handed component of the bulk field must have vanishing boundary conditions on each brane. Equation (\ref{eqn:fermioneom}) implies the following second-order differential equation for the left- and right-handed components:
\begin{equation}
\label{eqn:hardfermioneom}
\left(z^2\partial_z^2 + z^2 m_n^2 - \nu (\nu \mp 1)\right)f_n^{L,R} = 0.
\end{equation}
To begin with, we look for zero modes, $m_n=0$. Solving equation (\ref{eqn:fermioneom}) for the zero modes we get (for $\nu \ne \mp 1/2$),
\begin{equation}
f_0^{L,R}(z) = \frac{1}{L_1^{1/2}}\sqrt{\frac{1\pm2\nu}{1-\epsilon^{1\pm 2\nu}}} \left(\frac{z}{L_1}\right)^{\pm \nu}.
\end{equation}
For the case $\nu=-1/2$ ($\nu = 1/2$) for left- (right-) handed fermions the wave-function is $f_0 = z^{-1/2}/\sqrt{-\log(\epsilon)}$.

For the massive KK modes we need to solve equation (\ref{eqn:hardfermioneom}). For $\nu \ne \frac{1}{2} + N$ the solutions are simply Bessel functions
\begin{equation}
 f_n^{L,R}(z) = \sqrt{z}\left(A_n^{L,R} J_{\frac{1}{2} \mp \nu}(m_n z) +B_n^{L,R} Y_{\frac{1}{2} \mp \nu}(m_n z)\right)
\end{equation}
However, note that the left- and right-handed solutions are not independent because they are coupled through the first-order equation (\ref{eqn:fermioneom}).  The spectrum and normalization for the massive wave-function are found by imposing the different boundary conditions, together with equation (\ref{eqn:fermionnor}). From equation (\ref{eqn:fermioneom}) we see that if we choose Dirichlet boundary conditions for the right- (left-) handed mode we must choose the modified Neumann boundary condition $\left(z \partial_z \mp \nu\right)f_n^{L,R}(z) = 0$ for the left- (right-) handed modes. In the case of odd boundary conditions on both branes for a right-handed field, for instance, we recover the same constraint on $m_n$ as equation (\ref{eqn: KK mass spectrum}).

 \end{document}